\documentclass{jpp}

\usepackage{graphicx}
\usepackage{natbib}
\usepackage{pdflscape}
\usepackage{bm}
\usepackage{float}
\usepackage{array}
\usepackage{color,amssymb,hyperref}

\newcommand{\be}{\begin{equation}}
\newcommand{\ee}{\end{equation}}
\newcommand{\bea}{\begin{eqnarray}}
\newcommand{\eea}{\end{eqnarray}}

\newcommand{\GG}[1]{}
\newcommand{\Fig}[1]{Fig.~\ref{#1}}
\newcommand{\Tab}[1]{Table~\ref{#1}}
\newcommand{\eq}[1]{(\ref{#1})}

\newcommand{\Eq}[1]{Eq.~(\ref{#1})}
\newcommand{\Eqs}[2]{Eqs.~(\ref{#1}) and~(\ref{#2})}

\newcommand{\EEqs}[2]{Eqs.~(\ref{#1}) and~(\ref{#2})}

\def\delsp{\mbox{\rm $\delta$}_{SP}}
\def\delin{\mbox{\rm $\delta$}_{in}}
\def\rhos{\mbox{\rm $\rho$}_{s}}

\def\delp{\mbox{\rm $\Delta$}'}

\newcommand{\Bo}{B_{0}}
\newcommand{\noe}{n_{0e}}
\newcommand{\Toe}{T_{0e}}

\newcommand{\nabpsq}{\mbox{$\nabla^2_{\perp}$} {}}

\newcommand{\Apar}{A_{\parallel}}
\newcommand{\Tpare}{T_{\parallel e}}

\newcommand{\desq}{d^2_e}

\title[Plasmoid instability in the semi-collisional regime]{Plasmoid instability in the semi-collisional regime}

\author[Pallavi Bhat and Nuno F. Loureiro]
{P\ls A\ls L\ls L\ls A\ls V\ls I\ns B\ls H\ls A\ls T$^1$%
  \thanks{Email address for correspondence: pbhat@mit.edu},\break
\and\ns N\ls U\ls N\ls O\ns F.\ns L\ls O\ls U\ls R\ls E\ls I\ls R\ls O$^1$}

\affiliation{$^1$Plasma Science and Fusion Center, \\Massachusetts Institute of Technology, Cambridge, MA 02139, USA}

\date{\today}

\begin{document}

\maketitle

\begin{abstract}
We investigate analytically and numerically the semi-collisional regime of the plasmoid instability, defined by the inequality $\delsp \gg \rho_s \gg
\delin$, where $\delsp$ is the width of a Sweet-Parker current sheet, $\rhos$ is the ion sound Larmor radius, and $\delta_{in}$ is width of boundary
layer that arises in the plasmoid instability analysis.  Theoretically, this regime is predicted to exist if the Lundquist number $S$ and the length
of the current sheet $L$ are such that $(L/\rhos)^{14/9} < S <(L/\rhos)^2$ (for a sinusoidal-like magnetic configuration; for a Harris-type sheet the
lower bound is replaced with $(L/\rhos)^{8/5}$).  These bounds are validated numerically by means of simulations using a reduced gyrokinetic model
(Zocco \& Schekochihin, {\it Physics of Plasmas} {\bf 18}, 2011) conducted with the code {\tt Viriato}.  Importantly, this regime is conjectured to allow for plasmoid formation at
relatively low, experimentally accessible, values of the Lundquist number.  Our simulations obtain plasmoid instability at values of $S$ as low as
$\sim 250$.  The simulations do not prescribe a Sweet-Parker sheet; rather, one is formed self-consistenly during the nonlinear evolution of the
initial tearing mode configuration. This proves that this regime of the plasmoid instability is realizable, at least at the relatively low values of
the Lundquist number that are accessible to current dedicated experiments.
\end{abstract}

\section{Introduction}
Magnetic reconnection is a fundamental plasma physics phenomenon, relevant to  laboratory, space, and astrophysical systems \citep{biskamp_book,
ZY2009, uzdensky2011}.  It involves a rapid topological rearrangement of the magnetic field, leading to efficient magnetic energy conversion and
dissipation.  Solar flares \citep{shibata_solar_2011} and sawtooth crashes in tokamaks \citep{hastie_sawtooth_1997} are two popular examples of
processes where reconnection plays a key role; others include substorms in the Earth's magnetosphere \citep{dungey_1961,burch_16}, particle
acceleration in jets and pulsar winds \citep{ceruttietal12,Kagan2015, werner_2016}, magnetized turbulence
(e.g. \citet{matthaeus_86, servidio_09, zhdankinetal2013,loureiro_role_2017, mallet2017, cerri_17}), etc.

In magnetohydrodynamic (MHD) plasmas, reconnection sites (current sheets) tend to be unstable to the formation of multiple small islands (or
plasmoids) provided that the Lundquist number (defined as $S=L V_A/\eta$, where $L$ is the system size, $V_A$ the Alfv\'en speed, and $\eta$ the
magnetic diffusivity) is sufficiently large (typically, $S\gtrsim 10^4$).  This is known as the plasmoid instability; the current sheets mediated by
the plasmoids have an aspect ratio that is much smaller than that of the global sheet, thus triggering fast reconnection
\citep{nunoetal2007,lapenta_2008,BHY2009,cassak_2009, samtaney_2009, huang_2010,uzdenskyetal2010, loureiro_2012,loureiro_2013,nunouzdensky2016}. 

In weakly collisional plasmas, where the frozen-flux constraint is broken by kinetic effects instead of collisions, plasmoids are also abundantly
observed (e.g. \citet{drake_2006,JD2011,DR2012}), suggesting that plasmoid generation and dynamics are robust and fundamental features of reconnecting
systems, regardless of the collisionality of the ambient plasma.

Because of its perceived importance --- from determining the reconnection rate in MHD plasmas to its possible role on the reconnection
onset~\citep{pucci_2014, uzdensky_2016,comisso_2016,tolman2017} and in the energy partition~\citep{loureiro_2012, numata_2015} and particle
acceleration and plasma
heating~\citep{drake_2006,gianniaosetal2009,Okaetal2010,ceruttietal2013,sironispitkovsky2014,guoetal2015,zhouetal2015,sharmaetal2017,wernUzd2017} --- the plasmoid instability
has been the subject of a multitude of theoretical and numerical studies (see \citet{nunouzdensky2016} for a brief review).  There are also abundant
reports of plasmoid observation in solar flares \citep{nishizukaetal2010,milliganetal10,liuetal13}, coronal jets \citep{zhangJi14} and in the Earth's
magnetotail \citep{moldwinHughes92,zongetal2004} and in the magnetospheres of other planets \citep{jackmanetal2011,zhangetal12,Braccioetal15}.  This, however, contrasts
starkly with plasmoid detection and investigation in laboratory experiments, which has so far been relatively limited, with only a handful of studies
reporting plasmoid observation~\citep{foxetal2011,dorfman_2013,olson_2016,jara-almonte_2016,hare2017prl,hare2017xiv,hare_experimental_17}.  In all
cases, these observations have occurred in non-MHD regions of parameter space (dedicated reconnection experiments have not been able to reach
$S>10^4$, though future ones might~\citep{flare,trex}), and lack a solid theoretical footing.

In a recent paper, \cite{nunouzdensky2016} have identified a plasma collisionality regime where the requirement for triggering the plasmoid
instability is significantly eased with respect to its pure MHD counterpart.  In essence, this regime relies on collisionality being high enough that
an MHD current sheet may form in the first place (i.e., the current sheet thickness exceeds any kinetic scale); but small enough that when such a
sheet is analyzed for its stability to plasmoid formation, two fluid effects can no longer be neglected.
Interestingly, some of the above mentioned
experimental reports of plasmoid detection~\citep{dorfman_2013,jara-almonte_2016,hare2017prl,hare2017xiv,hare_experimental_17} seem to sit in, or very close to, this region of parameter
space~\citep{hare_thesis}, and it is conceivable that they provide experimental
evidence of the existence of this novel regime.

The aim of this paper is to report a set of numerical experiments designed to confirm the existence of the semi-collisional plasmoid
instability, with particular focus on experimentally acessible values of the Lunquist number, and precisely map out the regions of parameter space
inhabited by it.

\section{The semi-collisional plasmoid instability}
\label{sec:sc-plasmoid}

The linear theory of the plasmoid instability in MHD plasmas~\citep{nunoetal2007,BHY2009,loureiro_2013} assumes the existence of a Sweet-Parker (SP) 
current sheet~\citep{sweet_1958,parker_1957}, which is taken as the background equilibrium whose stability is analyzed.  In standard tearing mode
fashion~\citep{FKR}, the calculation divides the spatial domain into an outer region, where resistivity effects can be ignored, and an inner region
--- a boundary layer of thickness $\delta_{in}$ --- where resistivity matters.

Let us revisit this question adding minimal kinetic two-fluid effects: we wish to consider the case where the ion sound Larmor radius, $\rhos$, though
smaller than the thickness of the SP current sheet, $\delta_{SP}$, is however larger than the boundary layer of the MHD linear plasmoid
instability: $\delta_{SP} \gg \rho_s \gg \delta_{in}$~\footnote{An additional requirement is that the electron skin depth, $d_e = c/\omega_{pe}$, with
$\omega_{pe}$ the electron plasma frequency, is negligible, i.e., $\delta_{in}\gg d_e$. A further generalization of the theory to include electron
inertia effects is possible --- see ~\cite{nunouzdensky2016}.}.  
This obviously implies that MHD is no longer a sufficient description.
On the other hand, collisionality ($\nu_{ei}$) will still be taken to be large enough that the frozen flux constraint is broken by resistivity, 
instead of electron inertia; i.e., $\delta_{in}\gg d_e$, with $d_e$ the electron skin depth (which results from the ordering $\nu_{ei}\gg\gamma$, where $\gamma$ is the tearing mode growth rate).
Nonetheless, the semi-collisional tearing mode requires $\delta_{in}\ll \rho_s$ and transitions to the usual collisional (MHD) tearing mode in the limit $\delta_{in}\gg \rho_s$, 
and becomes the collisionless tearing mode when the collision frequency is decreased such that $\nu_{ei}\ll \gamma$ \citep{DL1977}.
On a qualitative level, this regime does not require ion finite Larmor orbit effects (i.e., it exists in the limit $\rho_i \rightarrow 0$ as long as $\rho_s$ remains finite). 
But our arguments below assume $\rho_i\sim \rho_s$, and our simulations will consider the case when $T_{0i}=T_{0e}$. 

The expressions for the growth rate of the plasmoid instability in this regime can be obtained from the appropriate tearing mode theory~\citep{DL1977,pegoraro1986,zs2011}, following the
usual procedure of replacing the equilibrium scale length with $\delta_{SP}\sim L S^{-1/2}$, where $L$ is 
the length of the current sheet \citep{tajimashibatabook,BHY2009}. 

For small values of the tearing mode instability parameter $\delp$, i.e., $\delp \delin \ll 1$, we find 
\bea
\label{gamma_smalldp}
\gamma L / V_A &\sim& (k L)^{2/3} (\delp \rho_s)^{2/3},\\
\delin/L &\sim& (\delp \rhos)^{1/6} (kL)^{-1/3} S^{-1/2}, 
\eea
where $\gamma$ is the growth rate of a mode with wavenumber $k$.
This expression can be simplified if $\delp$ is not too small, such that it can be approximated as $\delp \delta_{SP}\sim 1/(k\delta_{SP})$, 
as pertains to the usual Harris-like magnetic configuration \citep{harris62}.
In that case, we obtain
\be
\gamma L / V_A \sim (\rhos/L)^{2/3} S^{2/3},
\ee
and the validity condition $\delp \delin \ll 1$ becomes
$kL \gg (\rhos/L)^{1/9} S^{4/9}$.
Note that this expression is independent of $k$ to lowest order.

In the opposite limit of large $\delp$, i.e., $\delp \delin \gg 1$, we instead have
\bea
\label{largedp}
\gamma L/V_A &\sim& (\rhos/L)^{4/7} S^{2/7} (kL)^{6/7},\\
\delin/L &\sim& (\rhos/L)^{1/7} (kL)^{-2/7} S^{-3/7}.
\eea
The fastest growing mode is yielded by the intersection of these two branches \citep{nunouzdensky2016}\footnote{Note that the scaling for $\delin$, Eq.~\eq{delta}, 
has been corrected from~\cite{nunouzdensky2016}.} 
\bea
\label{gmax}
\gamma_{max} L/V_A &\sim& (\rho_s/L)^{2/3}S^{2/3}, \\
\label{kmax}
k_{max}L &\sim& (\rho_s/L)^{1/9} S^{4/9},\\
\label{delta}
\delin/L &\sim& (\rho_s/L)^{1/9}S^{-5/9}.
\eea

\begin{figure}
  \centerline{\includegraphics[height=6cm,width=8cm]{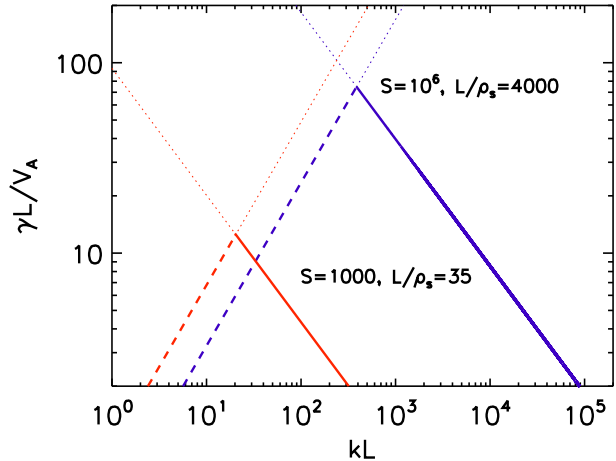}}
  \caption{Dispersion relation for the semi-collisional plasmoid instability for two different combinations of the relevant parameters: $(S,L/\rho_s)
= (1000,35)$ (red lines) and $(S,L/\rho_s) = (10^6,4000)$ (blue lines).  Solid lines represent the case of $\delp \delin \ll 1$ (\Eq{smallgam2}),
whereas dashed lines are for the case of $\delp \delin \gg 1$ (\Eq{largedp}). Respectively, these lines are only valid to the right, or to the left,
of their intersection. Their point of intersection provides an estimate of the most unstable wavenumber and corresponding growth rate.}
\label{growthrates}
\end{figure}
The validity of these expressions rests on two conditions: $\delta_{SP} \gg \rho_s$, and $\rho_s \gg \delin$.  Using Eq.~\eq{delta}, these therefore
imply that the semi-collisional plasmoid instability inhabits the region of parameter space defined by~\footnote{In addition, the existence of the plasmoid
instability (irrespective of the collisionality regime) requires that $\gamma L/V_A \gg 1$ and $k L \gg 1$; both of these conditions yield
requirements on $S$ and $L/\rho_s$ that are less demanding than the rightmost inequality in~\eq{SC_space}, so $(L/\rho_s)^{8/5}$ should be the correct
lower bound.}
\be
\label{SC_space}
(L/\rho_s)^2 \gg S \gg (L/\rho_s)^{8/5}.
\ee

An alternative current sheet profile worth considering -- and the one we will make use of in this paper -- is 
that of a sinusoidal-like magnetic field, {\bf $B_y(x)=B_0 \sin(x/a)$}, 
for which $\delp \delta_{SP}\sim 1/(k\delta_{SP})^2$.
For small $\delp$, we obtain,
\bea
\gamma L/V_A &\sim& (\rhos/L)^{2/3}(kL)^{-2/3} S,\\
\label{smallgam2}
\delin &\sim& (\rhos/L)^{1/6} (kL)^{-2/3} S^{-1/4},
\label{smalldelta2}
\eea
valid if $kL \gg (\rhos/L)^{1/16} S^{15/32}$.
For large $\delp$, the scaling for $\gamma L/V_A$ is same as in \Eq{largedp}, as there is no explicit dependence on $\delp$.
The fastest growing mode is therefore characterized by
\bea
\label{gam2}
\gamma_{max} L/V_A &\sim& (\rho_s/L)^{5/8}S^{11/16},\\
\label{wave2}
k_{max}L &\sim& (\rho_s/L)^{1/16} S^{15/32},\\
\label{delta2}
\delta_{in}/L &\sim& (\rho_s/L)^{1/8}S^{-9/16}.
\eea
\Fig{growthrates} illustrates both limits of the dispersion relation, and their intersection, for two different combinations of the two relevant
parameters, $S$ and $L/\rho_s$.
In appendix~\ref{validation} we recover these scalings via direct numerical simulation, confirming both their validity and the ability of the code
{\tt Viriato}~\citep{nunoetal2016}, which we will employ in this paper (see Sections~\ref{model} and \ref{setup}), to recover them.

In this case of sinusoidal-like current sheet profile, Eq.~\eq{SC_space} is replaced by: \be (L/\rho_s)^2 \gg S \gg (L/\rho_s)^{14/9}. \label{SC2} \ee
The lower bound here has only a slightly smaller exponent than (and in practice difficult to discern from) the Harris-like case of \Eq{SC_space}.

\Eqs{SC_space}{SC2} lead to the interesting suggestion that this particular version of the plasmoid instability can be obtained at relatively low
values of $S$, provided that the system (the current sheet length $L$, to be precise) is not too large compared with $\rho_s$. In other words, the
lower bound of $S\sim 10^4$ that pertains to the MHD version of the plasmoid instability~\citep{biskamp86,nunoetal2005,samtaney_2009,baty_2014} is
replaced by a {\it function} in the semi-collisional regime, $(L/\rhos)^{14/9}$, or $(L/\rhos)^{8/5}$, as appropriate.  
This growth of plasmoids at low $S$ is possible in the semi-collisional regime due to the presence of the $\rhos$ scale, which 
is unavailable in MHD. The tearing calculation in this regime has to account for nested boundary layers instead of a single one: this introduces 
an extra parameter in the problem and makes the critical Lundquist number (derived from the requirement that the condition for semi-collisional 
regime, $\delta_{SP}>\rho_s>\delta_{in}$ be satisfied) become dependent on this new parameter.

From both the experimental and
the numerical points of view, this feature of low critical Lundquist number is a significant advantage.  In particular, this regime should be available to reconnection experiments such as
Magnetic Reconnection Experiment (MRX) \citep{yamadaetal1997}, Terrestrial Reconnection Experiment (TREX) \citep{olsonetal2016}, Facility for Laboratory Reconnection Experiments (FLARE) 
\citep{JD2011} and Mega Ampere Generator for Plasma Implosion Experiments (Magpie)~\citep{hare2017prl,hare2017xiv,hare_experimental_17}. Indeed, as we mention above, it is tempting
to attribute recent reports of experimental plasmoid observation~\citep{dorfman_2013,jara-almonte_2016,hare2017prl,hare2017xiv,hare_experimental_17}
to this version of the plasmoid instability --- or to its $\beta\sim1$ analogue~\citep{baalrud_11} ($\beta$ is the ratio of the plasma pressure to the
magnetic pressure). In the weak guide field or $\beta\sim1$ case, the relevant scale is ion-inertial scale $d_i$ (instead of $\rhos$), and the same feature 
of scale-dependent critical Lundquist number is expected to hold. 

Despite these speculations and conjectures, the existence of this regime of the plasmoid instability has not been confirmed via direct numerical
simulations, and indeed there are a couple of issues that may raise suspicion.  In particular, we note that all of the scalings above are predicated
on there being an asymptotic separation between the scales involved, namely $\delta_{SP} \gg \rho_s \gg \delin$ \footnote{\label{footnote} Note that the condition for
semi-collisional regime should not be confused with a similar looking criterion specified in a scenario of hierarchy of plasmoids and interplasmoid
current sheets, namely, $\delta_{SP}(L) > \rho_s > \delta_c$ (see \cite{uzdenskyetal2010}).  Here, $\delta_{SP}(L)$ refers to the width of the primary
global SP current sheet and $\delta_c$ is the width of the smallest interplasmoid SP current sheet which is marginally unstable to the plasmoid
instability.  This criterion guarantees a transition of the
plasmoid hierarchy from MHD to kinetic scales.\\

The scalings \Eqs{delta}{delta2} allow us to make this criterion more precise.
Consider the plasmoid cascade.
For an interplasmoid current sheet with the width $\delta^{(N)}$ ($\delta^{(2)}$, $\delta^{(3)}$ are the secondary and tertiary SP current sheets, and so on), if the
corresponding inner layer $\delta^{(N)}_{in}$ is larger than $\rhos$, then the arising plasmoids are still in MHD regime.
However, if at any point in the plasmoid cascade one obtains $\rhos >
\delta^{(N)}_{in}$,
then the system transitions into the
semi-collisional regime.
Using the relationship $\delta_c \sim L_c S_c^{-1/2}$, where $S_c\sim 10^4$ is the critical value of the Lundquist number to obtain the plasmoid instability in MHD, 
it is easy to conclude that a transition to the semi-collisional regime must occur at some point in the plasmoid hierachical cascade if $S>S_c^{3/8} (L/\rho_s)$, 
for a Harris-type sheet, or $S>S_c^{5/14}(L/\rho_s)$, for a sinusoidal-type sheet.}.
As the Lundquist number is made smaller, this scale separation is inevitably lost, and thus the claim that the
semi-collisional plasmoid instability may be obtainable at the relatively low values of $S$ and $L/\rho_s$ that are within the reach of existing
experiments needs careful numerical validation.

An additional concern of significant relevance is that of whether the Sweet-Parker sheet that is assumed as the background for the instability derived
here is realizable.  As has been pointed out~\citep{nunoetal2007,pucci_2014,uzdensky_2016}, the existence of a super-Alfv\'enic instability whose
growth rate diverges as $S\rightarrow \infty$ indicates that the equilibrium that it arises from, may never form in the first place.  We think this
claim is pertinent at high values of the Lundquist number.  However, in the opposite limit of relatively low $S$ with which we are mostly concerned
here, where the instability is only mildly super-Alfv\'enic, a Sweet-Parker sheet may still form and beget the instability.

This paper aims to answer these questions by means of direct numerical simulations of a reconnecting system where the current sheet is not prescribed,
but rather allowed to form self-consistently. 

\section{\label{model}The model}
The weak collisionality of a large number of reconnecting environments demands the use of a kinetic description.  In the most general case, one is
forced to adopt a first-principles formalism (such as particle-in-cell, or the 6D Vlasov (or Boltzmann) equation), with its inherent analytical and
numerical complexity.  In many situations, however, a strong component of the magnetic field is present that is perpendicular to the reconnection
plane.  This guide-field offers an opportunity for significant simplification: the 5D gyrokinetic formalism \citep{FC1982,howesetal2006}. 

Further simplification is possible if, in addition, one considers plasmas such that the electron beta is sufficiently low to be comparable to the 
electron-to-ion mass ratio ($m_e/m_i$); a case in point is the solar corona, as mentioned earlier in section \ref{sec:sc-plasmoid}.  Leveraging on
these assumptions (strong guide-field and low beta), a reduced-gyrokinetic formalism was recently derived (dubbed `Kinetic Reduced Electron Heating
Model', or KREHM) \citep{zs2011}.  One of its appealing features is that the phase-space is reduced further to 4D (position vector and velocity in the
guide-field direction only), thus rendering computations, and even analytic theory, significantly more manageable than fully kinetic approaches.

In this work, we use the KREHM equations to investigate the plasmoid instability in the semi-collisional regime.  We will restrict our numerical
investigations to the two spatial dimensions comprising the reconnection plane, $(x,y)$.
With this restriction, the KREHM equations become: \be
\frac{1}{\noe}\frac{d\delta n_e}{dt} = \frac{1}{B_0} \left\{\Apar, \frac{e}{c m_e}\desq \nabpsq
\Apar \right\}, \label{pertden} \ee \be
\frac{d}{dt} (\Apar- \desq \nabpsq \Apar) = ~\eta \nabpsq \Apar  
- \frac{c\Toe}{eB_0} \left\{ \Apar, \left( \frac{\delta n_e}{\noe} + \frac{\delta \Tpare}{\Toe}
  \right) \right\}, \label{current} \ee
\be \frac{dg_e}{dt} - \frac{v_{\parallel}}{B_0} \left\{\Apar,\left(g_e-\frac{\delta
\Tpare}{\Toe}F_{0e} \right) \right\} = C[g_e] - \left(1-\frac{2v_{\parallel}^2}{v_{the}^2}\right)
\frac{F_{0e}}{B_0} \left\{\Apar, \frac{e}{c m_e}\desq \nabpsq \Apar\right\},  \label{edist}
\ee where \be \frac{d}{dt}=\frac{\partial}{\partial t} + \frac{c}{\Bo} \{\phi, \dots\}.  \ee
Here, $\phi$ is electrostatic potential and $\{\dots\}$ denotes the Poisson bracket, defined for any two fields $P$, $Q$ as $\{P, Q \} = \partial_xP
\partial_yQ -\partial_y P\partial_x Q$.  

\EEqs{pertden}{current} evolve the zeroth and first moments of the perturbed electron distribution function, $\delta f_e$.  The zeroth moment is the electron density perturbation, $\delta n_e$. 
The first moment is the parallel electron flow, $u_{\parallel e}$, and is related to the parallel component of the magnetic vector potential $\Apar$ by $u_{\parallel e}=(e/cm_e)\desq \nabpsq \Apar$.
The perturbed electron distribution function is given by \be \delta f_e = g_e+ ({\delta n_e}/{\noe}+2v_{\parallel}u_{\parallel e}/v_{the}^2)F_{0e}, \ee 
where $F_{0e}= \noe/(\pi/v_{the}^2)^{3/2}$exp$[-(v_{\parallel}^2 + v_{\perp}^2 )/v_{the}^2]$ is the Maxwellian equilibrium, and $v_{\parallel}$ and
$v_{\perp}$ are, respectively, the velocity coordinates parallel and perpendicular to the guide-field direction.
The electron thermal speed is $v_{the} =\sqrt{2T_{0e}/m_e}$, with $e$ the electron charge, and $m_e$ its mass.
Note that this is a $\delta f$ formulation, so any fluctuating quantity is necessarily much smaller than the equilibrium quantity. 

The quantity $g_e$ is dubbed the reduced electron distribution function; its evolution is given by \Eq{edist}. 
It contains all the moments of $\delta f_e$ higher than $\delta n_e$ and $u_{\parallel e}$.
For example, parallel temperature fluctuations (second order moment) are given by $\delta \Tpare/\Toe =
(1/\noe)\int d^3{\bf v} (2v_{\parallel}^2/v_{the}^2)g_e$.
On the right-hand side, $C[g_e]$ denotes the collision operator, and the second term represents what survives of the so-called ``gyrokinetic potential'' in this expansion.

The background magnetic guide field is $\Bo$; and $\noe$, $\Toe$ are the background electron density and
temperature, respectively.
Other symbols introduced above are the electron skin depth, $d_e=c/\omega_{pe}$, with $\omega_{pe}=\sqrt{4\pi \noe
e^2/m_e}$ the electron plasma frequency.

The perturbed electron density, $\delta n_e$, and the electrostatic potential are related via the gyrokinetic Poisson law,
\be
\frac{\delta n_e}{n_{0e}} = \frac{1}{\tau} (\hat{\Gamma}_0 -1)
\frac{e\phi}{T_{0e}}.
\label{ne}
\ee
where, $\tau=T_{0i}/\Toe$ is the ion to electron background temperature ratio, and $\hat \Gamma_0$ is a real space operator whose Fourier transform is
$\Gamma_0(\alpha)=I_0(\alpha) e^{-\alpha}$, with $I_0$ the modified Bessel function of zeroth order, and $\alpha={k^2_{\perp}} {\rho_i}^2/2$, with
$\rho_i=v_{thi}/\Omega_i$ the ion Larmor radius,  $v_{thi}=\sqrt{2T_{0i}/m_i}$ the ion thermal velocity, and $\Omega_i=|e|\Bo/m_ic$ the ion gyro
frequency.

To make contact with a more familiar set of equations, note that, in the fluid limit ($d_e,~\rho_i, ~\rho_s \rightarrow 0$), \EEqs{pertden}{current}  decouple from \Eq{edist} 
and reduce to the momentum and induction equations of reduced MHD. 
However, when kinetic effects are retained, Ohm's law couples to the kinetic equation (\Eq{edist}) 
via parallel electron temperature fluctuations.

Observe that \Eq{edist} does not explicitly depend on $v_{\perp}$.  Provided that one chooses a collision operator which is itself also independent of
$v_\perp$, this coordinate can be integrated out, yielding a reduced electron distribution function that is effectively four dimensional only,
$g_e=g_e(x,y,z,v_{\parallel},t)$.  One model collision operator that exhibits this property is the Lenard-Bernstein operator \citep{LenBern}, modified
to conserve the pertinent quantities~\citep{zs2011}.
Then \Eq{edist} can be more conveniently solved in terms of its expansion in Hermite polynomials: \be
g_e(x,y,z,v_{\parallel},t)=\sum_{m=2}^\infty\frac{1}{2^m m!} H_m(v_{\parallel}/v_{the}) g_m(x,y,z,t) F_{0e}(v_{\parallel}), \ee where $H_m$ denotes
the Hermite polynomial of order $m$ and $g_m$ is its coefficient.  Inserting this expansion into \Eq{edist} yields a coupled set of fluid-like
equations for the coefficients of the Hermite polynomials, where by construction $g_0=g_1=0$, and for $m\ge2$ we have  
\bea 
\frac{dg_m}{dt} &-& \frac{v_{the}}{B_0}~
\left( \sqrt{\frac{m+1}{2}} \left\{\Apar, g_{m+1}\right\} + \sqrt{\frac{m}{2}} \left\{\Apar,
g_{m-1}\right\} \right) \nonumber\\ &=&  \frac{\sqrt{2} \delta_{m,2}}{B_0}~ \left\{\Apar,
\frac{e}{cm_e}d_e^2 \nabpsq \Apar\right\} - \nu_{ei} (mg_m-2\delta_{m,2}g_2), \label{gherm} 
\eea
where $\nu_{ei}$ is the electron-ion collision frequency, $\nu_{ei} = \eta/d_e^2$.
The kinetic equations solved by means of Hermite expansion requires a closure. A way to close the set of equations is to demand that 
at some $m=M$, the collision term becomes significant such that $g_{M+1}/g_{M}\ll 1$. This constraint will truncate the kinetic equations at $g_M$ 
as $g_{M+1}$ can be written in terms of $g_M$ \citep{zs2011,zocco_2015,nunoetal2016,whitehazeltine17}. This type of closure also recovers the semi-collisional limit exactly.

\section{Numerical setup}
\label{setup}
 
Eqs. (\ref{pertden}), (\ref{current}), (\ref{ne}) and (\ref{gherm}) are solved numerically on a two-dimensional grid of size $L_x \times L_y$, using
the pseudo-spectral code {\tt Viriato} \citep{nunoetal2016}.  Periodic boundary conditions are employed in both directions.  The numerical
configuration is akin to that employed in \cite{nunoetal2005} and, as we will show, is such that one can self-consistently obtain an SP current sheet
whose stability to plasmoid formation can then be studied.  Specifically, the input parameters can be specified in a way as to lead to the dynamic
formation of a SP current sheet that meets the conditions of the semi-collisional regime that we have discussed earlier.

The initial equilibrium is $\Apar(x,y,t=0)=A_{\parallel 0}/\cosh^2(x)$, where $A_{\parallel 0}=3\sqrt{3}/4$, such that the maximum of the reconnecting
field, $B_y=d\Apar/dx$, is $B_{y, max}=1$.  This equilibrium is destabilized with a small amplitude (linear) perturbation which seeds the fastest
growing tearing mode; in all simulations, this is the longest wavelength mode that fits in the simulation box.  Once in the nonlinear stage, the
tearing mode undergoes $X$-point collapse \citep{waelbroeck89, nunoetal2005}, and a current sheet forms which is consistent with the SP scaling (as we
shall confirm).  The plasmoid instability is then triggered, or not, depending on the values of $S$ and $\rho_s$ specified in the simulation.

The length of the SP current sheet can be varied by changing the instability parameter, $\Delta'(k)$, pertaining to the initial tearing
mode~\citep{nunoetal2005}.  In practice, this is achieved by changing the length of the box in the outflow direction, $L_y$. 

The resolution of a simulation in the inflow ($x$) direction is set by the size of the inner boundary layer that is expected to arise due to the
plasmoid instability, estimated using \Eq{delta}.  The resolution demands in the outflow ($y$) direction are less stringent, and are determined on a
{\it ad hoc} basis.

An additional constraint on our runs is that the electron inertia play no role. This is insured by setting it to be smaller than the resolution for
any given simulation. Therefore, in all runs, the frozen-flux constraint is broken by resistivity.  Note that no hyper-resistivity (or
hyper-viscosity) is used in these runs, ensuring that the actual Lundquist number in the simulations is determined by the resistivity that we specify.
The simulations also employ a finite viscosity, set equal to the magnetic diffusivity. 

A final choice has to do with the number of Hermite polynomials to keep in the simulations. For all runs reported here, the highest order polynomial
is  $M=4$ --- this ought to be sufficient given the relatively high collisionality of our simulations (and indeed we find that in all our runs
resistivity is the dominant energy dissipation channel).  To check convergence, we performed one test using instead $M=10$, and observed that in the
spectrum of $\vert g_m\vert^2/2$, the energy at $m>3$--$4$, is lower than that at $m=2$ by orders of magnitude, indicating that the power transferred
to the Hermite polynomials of higher order is not significant.  In all runs, the convergence of the Hermite representation is accelerated by the use
of a hyper-collision operator (see~\cite{nunoetal2016} for details).
    
\section{Results}
\begin{table}
\begin{center}
\caption{
Summary of all the runs discussed in the paper. 
The table lists the values of the sound Larmor radius $\rhos$, 
length of the current sheet $L$, their ratio $L/\rhos$, Lundquist number $S$, the inner boundary layer width $\delta_{in}$ 
(using \Eq{delta2}), the number of grid points employed $N_x \times N_y$, the length of domain in the $y$-direction, $L_y$, resistivity $\eta$, and answers whether 
both constraints of the semi-collisional regime are satisfied or not for a given run.}
\vspace{10pt}{\begin{tabular}{lcccccccccc}
\hline
\hline
Run & $\rhos$ & $L$ & $L/\rhos$ & $S$ & $\delin$ & $N_x \times N_y$ & $L_y/2\pi$ & $\eta$ & $\delsp > \rhos$? & $\rhos > \delin$?\\
\hline
A & 0.12  & 3.01 & 25.1 & 256    &  0.089  &  512 $\times$ 128  & 1.75  & 0.012   & Yes & Yes \\ 
B & 0.1   & 2.73 & 27.3 & 453    &  0.058  &  1024 $\times$ 256 & 1.695 & 0.00657 & Yes & Yes \\ 
C & 0.07  & 2.91 & 41.6 & 557    &  0.052  &  512 $\times$ 128  & 1.678 & 0.00583 & Yes & Yes \\ 
D & 0.06  & 2.58 & 43.0 & 977    &  0.033  &  1024 $\times$ 256 & 1.678 & 0.003   & Yes & Yes \\ 
E & 0.036 & 2.52 & 70.0 & 1149   &  0.028  &  1024 $\times$ 256 & 1.587 & 0.0025  & Yes & Yes \\ 
F & 0.04  & 2.28 & 57.0 & 1796   &  0.020  &  1024 $\times$ 256 & 1.587 & 0.00146 & Yes & Yes \\ 
G & 0.03  & 2.30 & 76.7 & 2997   &  0.015  &  2048 $\times$ 512 & 1.587 & 0.00088 & Yes & Yes \\ 
H & 0.015 & 2.46 & 164.0 & 5560   &  0.010  &  4096 $\times$ 512 & 1.587 & 0.0005 & Yes & Yes \\ 
I & 0.02  & 0.98 & 49.0 & 269	 &  0.026  &  512 $\times$ 128  & 1.19  & 0.004   & Yes & No  \\ 
K & 0.01  & 0.95 & 95.0 & 1092   &  0.010  &  1024 $\times$ 256 & 1.09  & 0.001   & Yes & No  \\ 
J & 0.02  & 1.56 & 78.0 & 565    &  0.026  &  1024 $\times$ 256 & 1.351 & 0.00276 & Yes & No  \\ 
L & 0.083 & 1.00 & 12.0 & 255	 &  0.032  &  512 $\times$ 128  & 1.19  & 0.004   & No  & Yes \\ 
M & 0.12  & 1.90 & 15.8 & 472    &  0.042  &  512 $\times$ 128  & 1.492 & 0.0044  & No  & Yes \\ 
N & 0.04  & 1.10 & 27.5 & 1171   &  0.014  &  1024 $\times$ 256 & 1.19  & 0.00094 & No  & Yes \\ 
\hline
\label{runs}\end{tabular}}
\end{center}
\end{table}
As previously stated, our main aim is to numerically ascertain the existence of the semi-collisional plasmoid instability  and validate the bounds of
the parameter space defined by $S$ and $L/\rho_s$ where this instability is expected to be active.
We specifically wish to focus on the instability's existence at modest, experimentally accessible, values of the Lundquist number.  To this effect, we
perform a series of runs as listed in~\Tab{runs}.  Amongst other parameters, the table lists the length of the current sheet, $L$, that is dynamically
obtained during the nonlinear evolution of the primary tearing mode (which results from the collapse of the $X$-point, as previously described).  This
length is measured using a full width at half maximum estimate\footnote{This is different from \citet{JD2011}, where $L$ represents half length of the reconnecting sheet. 
This difference amounts to a shift equal to $\log{(2)}$ between our reconnection phase diagram and theirs.} 
(as is the current sheet thickness, $\delta$) just before the plasmoid appears, and it is this length that is used to
estimate the Lundquist number that is also listed in ~\Tab{runs} (the magnitude of the upstream magnetic field remains unchanged by the $X$-point
collapse).

The first step in the description of our results is the characterization of the current sheet that is dynamically obtained from the $X$-point collapse
of the primary tearing mode.  The theory of the semi-collisional plasmoid instability laid out in Section~\ref{sec:sc-plasmoid} assumes a SP 
sheet as the background equilibrium; and so it is important to determine whether indeed that is the case in our simulations.
\begin{figure}
  \centerline{\includegraphics[height=6cm,width=8cm]{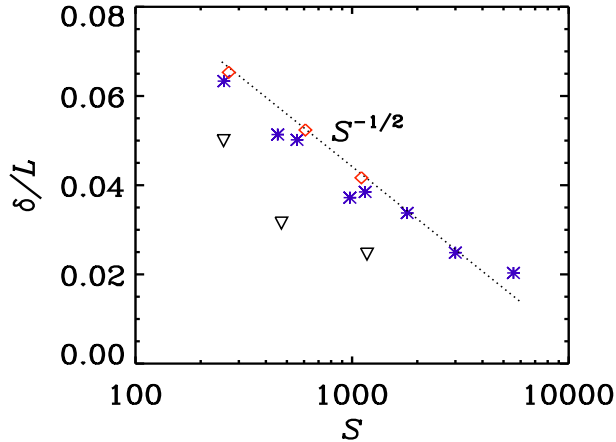}}
  \caption{ 
The ratio of current sheet width $\delta$ to its length $L$ against the Lundquist number $S$ is shown
for Runs~A through H (blue stars), I, J and K (red diamonds), and L, M and N (black inverted triangles).
Refer to Table~\ref{runs} for details of
these runs.
The dotted line indicates an $S^{-1/2}$ slope, expected for Sweet-Parker current sheets.}
\label{cs}
\end{figure}
In \Fig{cs}, we plot $\delta/L$ as a function of the Lundquist number $S$, obtained from all the runs.  
We find, as shown in \Fig{cs} by the blue
stars (Runs~A to H) and red diamonds (Runs~I, J, K), that the current sheets in these runs follow the SP scaling.  The system in these runs is
initially purely in the MHD regime as the inner boundary layer thickness of the primary tearing mode is larger than the kinetic scales.  And thus,
upon $X$-point collapse of the MHD tearing mode, the current sheets that form are expected to follow the SP scaling~\citep{nunoetal2005}, which indeed
bears out.

In the case of the three black inverted triangles shown in the \Fig{cs}, corresponding to Runs~L, M and N, the ion sound Larmor radius $\rhos$ is set
to be larger than the SP current sheet width ($\delsp$). 
Thus, when $X$-point collapse happens, these runs are affected by ion scale physics; unsurprisingly, the current sheets here do not follow the SP scaling. 
In summary, by the end of the collapse of the primary tearing mode and formation of the current sheet, 
the blue starred runs (Runs~A to H) are in the semi-collisional regime ($\delsp > \rhos > \delin$), the red diamond runs (Runs~I, J, K) in the MHD regime ($\delsp > \delin > \rhos$), and the 
black inverted triangles (Runs~L, M and N) in a weakly collisional regime where a Sweet-Parker sheet no longer forms ($\rhos > \delsp > \delin$).

\begin{figure}
  \centerline{\includegraphics[height=9cm,width=12cm]{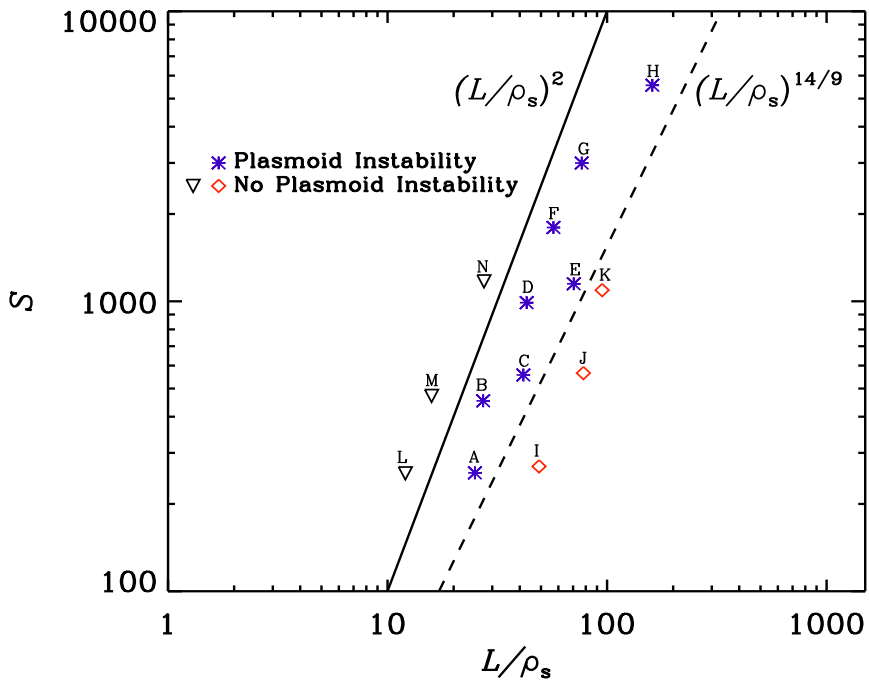}}
  \caption{Reconnection phase diagram showing the semi-collisional regime marked out by the upper bound ($\delta_{SP}>\rho_s$; solid line) and lower bound ($\rho_s>\delta_{in}$; dashed line)
	as prescribed by \Eq{SC2}.
    Blue stars denote simulations which show the plasmoid instability; red diamonds and black inverted triangles correspond to simulations which do not.
    The parameters (and other details) of each run can be looked up in Table~\ref{runs}.
}
\label{scbounds}
\end{figure}
Next, we show whether or not the plasmoid instability is observed for a given run on the $S$-$L/\rhos$ parameter space in \Fig{scbounds}.  The two
groups of runs represented by blue stars and red diamonds, which form SP current sheets, have $\rho_s<\delsp$ (the symbols here match those in 
\Fig{cs}).  Furthermore, in the case of the blue starred runs, upon the formation of the current sheet, the system becomes sensitive to
the presence of $\rhos$, as this kinetic scale is now larger than the semi-collisional inner boundary layer $\delin$ (corresponding to the newly formed SP layer) (see \Eq{delta}).  As a result,
the plasmoid instability can arise, and indeed it does, as seen in Runs~A to H (blue stars). On the other hand, in the case of Runs~I, J, K (red diamonds),
$\rhos$ is not only smaller that $\delsp$, but is also smaller that $\delin$ and thus no plasmoids arise in these runs.  The condition of $\rhos <
\delin$ implies the absence of the plasmoid instability simply because the system continues to be in MHD regime and $S$ is much below the critical
value of $\sim 10^4$ required in MHD. 

The group of runs A to H which do result in plasmoid instability are within the two theoretical bounds of the semi-collisional regime marked by solid
and dashed black lines (\Eq{SC2}).
Runs~I, J and K  (where $\rhos < \delin$) below the lower bound match roughly in $S$ with Runs~A, C and E respectively; Runs~L, M, N (where $\rhos >
\delsp$) above the upper bound, were performed to match in $S$ roughly with Runs~A, B and E respectively.  We find that neither of these two sets of
runs yield the plasmoid instability.  We conclude, therefore, that the theoretically prescribed bounds demarcating the semi-collisional regime are
remarkably robust within the range of Lundquist numbers and Larmor radii that we have explored. 
\begin{figure}
  \centerline{\includegraphics[height=4cm,width=14cm]{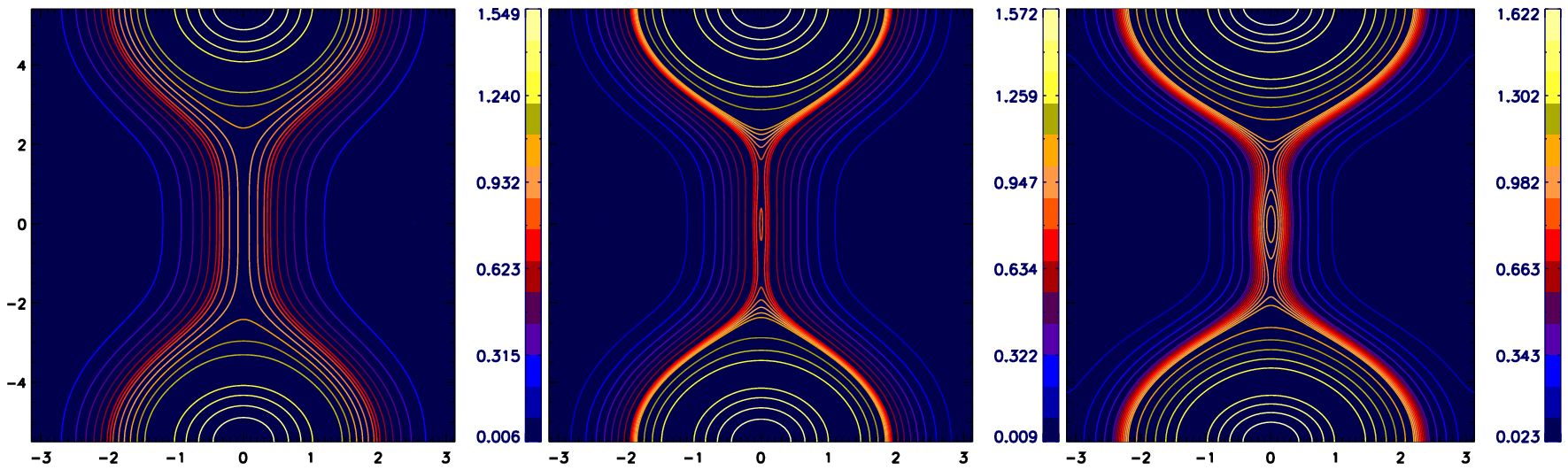}} 
  \centerline{\includegraphics[height=4cm,width=14cm]{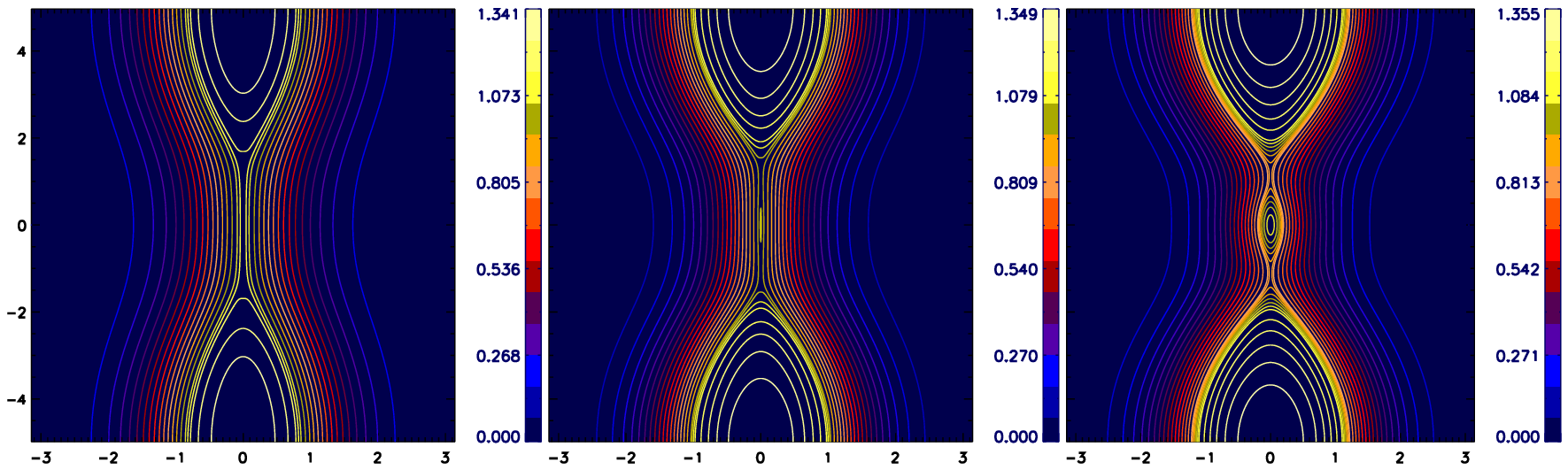}} 
  \caption{Contour plots of $A_\parallel$ for Run~A (top row) with lowest $S=256$ (and $L/\rhos=25$), 
and Run~H (bottom row) with highest $S=5560$ (and $L/\rhos=164$) at three different times: 
left panel, just before the plasmoid forms; middle panel, in the early stages of plasmoid formation; and right panel, when the plasmoid is in its nonlinear stage.} 
\label{contour1} 
\end{figure} 

Further insight can be gained by visually analyzing these runs.  In particular it is of interest to visually compare runs A and H, corresponding,
respectively, to the lowest ($S=256$) and highest ($S=5560$) values of the Lundquist number at which we have observed the plasmoid instability.
\Fig{contour1} shows contour plots of $\Apar$ at three different times.  The two leftmost panels (top and bottom) depict the system just before the
plasmoid formation.  The middle panels show the early stages of the plasmoid development; and the rightmost panels show the plasmoid well into the
nonlinear stage.  In both cases, note that in its early stage (middle panel) the $y$-extent of the plasmoid (roughly its linear wavelength) is much
smaller than the length of the current sheet (to be discussed below).  Also, due to the highly symmetric configuration of the magnetic field (and the
intrinsic symmetry germane to the pseudo-spectral method that we employ), the plasmoid is stuck to the middle of the sheet.  In a less constrained
situation, we expect that this plasmoid would be ejected upwards or downwards, and subsequent plasmoids to be seeded until the system approaches
saturation. It can be seen clearly that the gradient of $\Apar$ in the current sheet is larger in Run H compared to Run A, as expected given the order of 
magnitude difference in the values of their Lundquist numbers. Thus a curiosity is that even in the run with highest $S$, only a single plasmoid forms. We will address this concern at a later point.

\begin{figure}
  \centerline{\includegraphics[height=4cm,width=14cm]{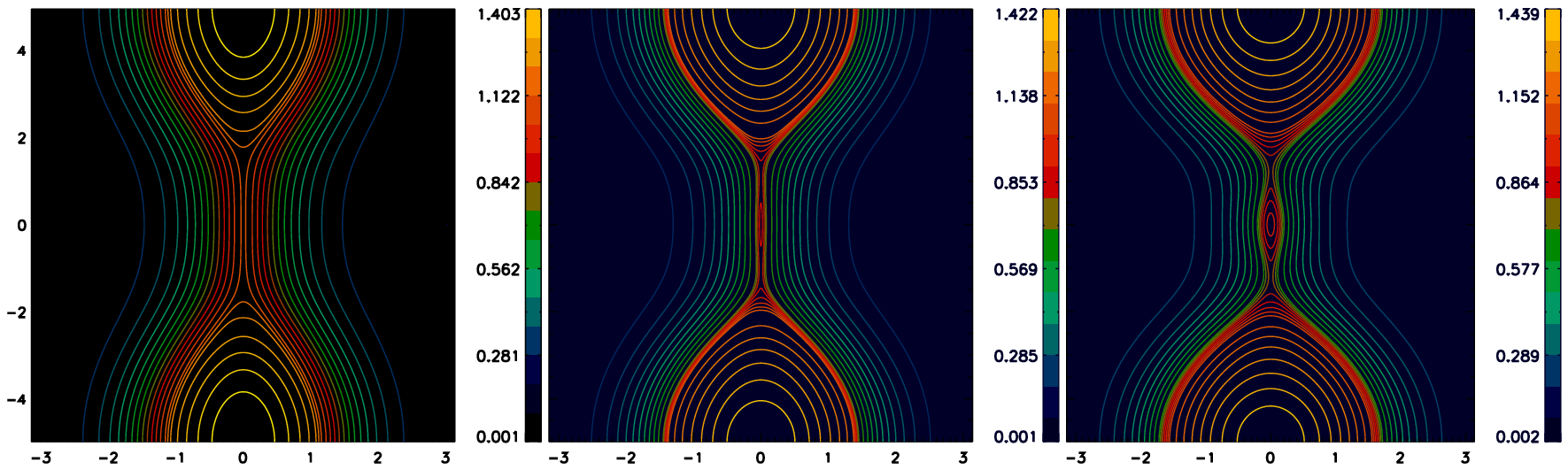}}
  \centerline{\includegraphics[height=4cm,width=14cm]{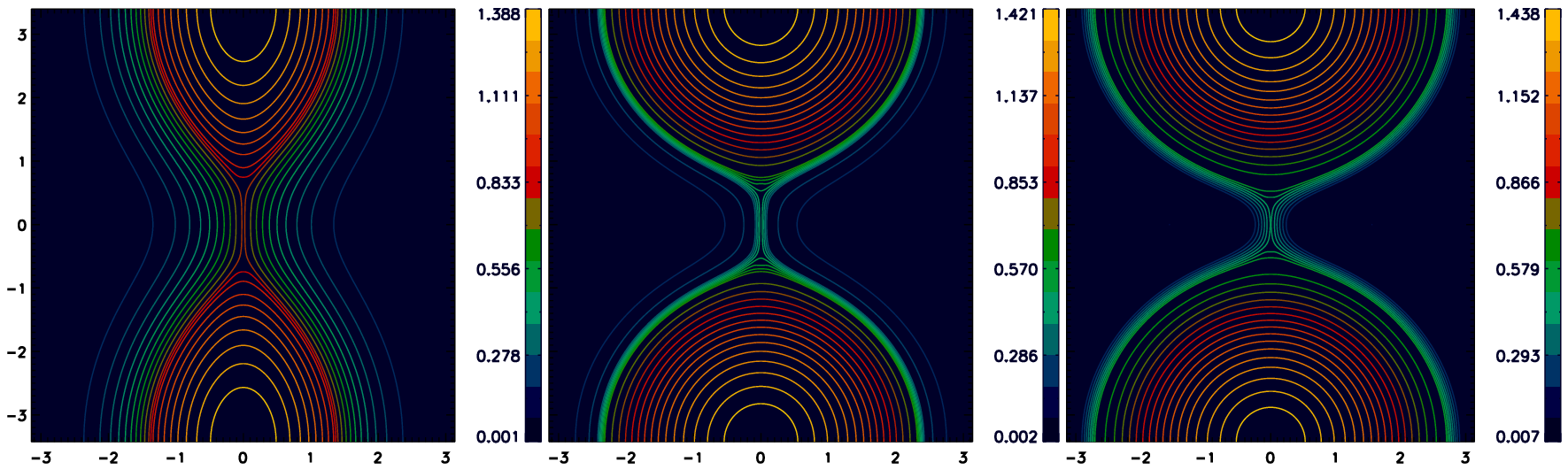}}
  \caption{Contour plots of $A_\parallel$ for Run~E in the semi-collisional regime (top row), with $S=1149$ and $L/\rhos=70.5$; and for Run~K in the MHD regime (bottom row), with $S=1092$ and $L/\rhos=95$, at three consecutive times from left to right.}
  \label{contour3}
\end{figure}

Another interesting comparison is between Runs~E ($S=1149$, $L/\rho_s=70.5$, blue star) and K ($S=1092$, $L/\rho_s=95$, red diamond).
As seen in~\Fig{contour3}, Run~E
shows a similar time evolution to that displayed by the runs in Fig.~\ref{contour1}, with the current sheet becoming unstable to plasmoid formation.
In Run K, on the other hand, the collapsed current sheet never goes unstable, and the primary tearing mode just proceeds to saturation.  This is
rather remarkable given how close in the phase-space outlined by $S$ and $L/\rho_s$ these two runs are (see Fig.~\ref{scbounds}).

\begin{figure}
  \centerline{\includegraphics[height=4cm,width=14cm]{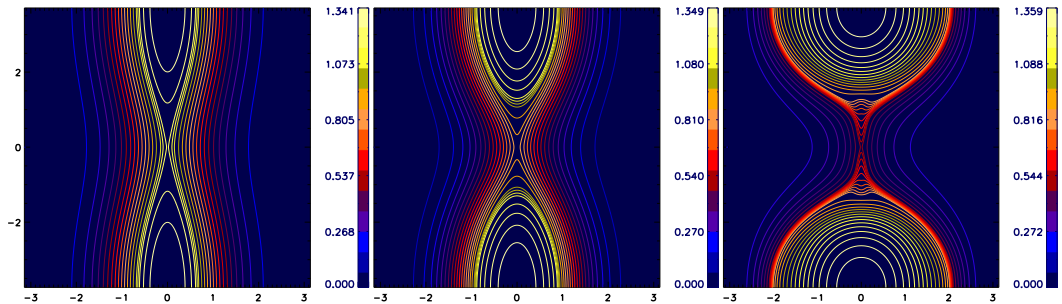}}
  \caption{Contour plots of $A_\parallel$ for Run~N (in the weakly collisional regime), with $S=1171$ and $L/\rhos=27.5$, at three consecutive times from left to right.}
  \label{contour4}
\end{figure}
In \Fig{contour4}, we also show a run from the weakly collisional regime, with the parameters $S=1171$ and $L/\rhos=27.5$ (Run~N). 
There is no plasmoid formation as expected. It is interesting to note that the current density forms an $X$-point, 
as seen from the leftmost panel in the \Fig{contour4}, which is consistent with what is expected in such a regime. 
The $X$-point later transitions to a double-structured current sheet, which retains a sharp peak at its center.

\begin{figure}
  \centerline{\includegraphics[height=6cm,width=8cm]{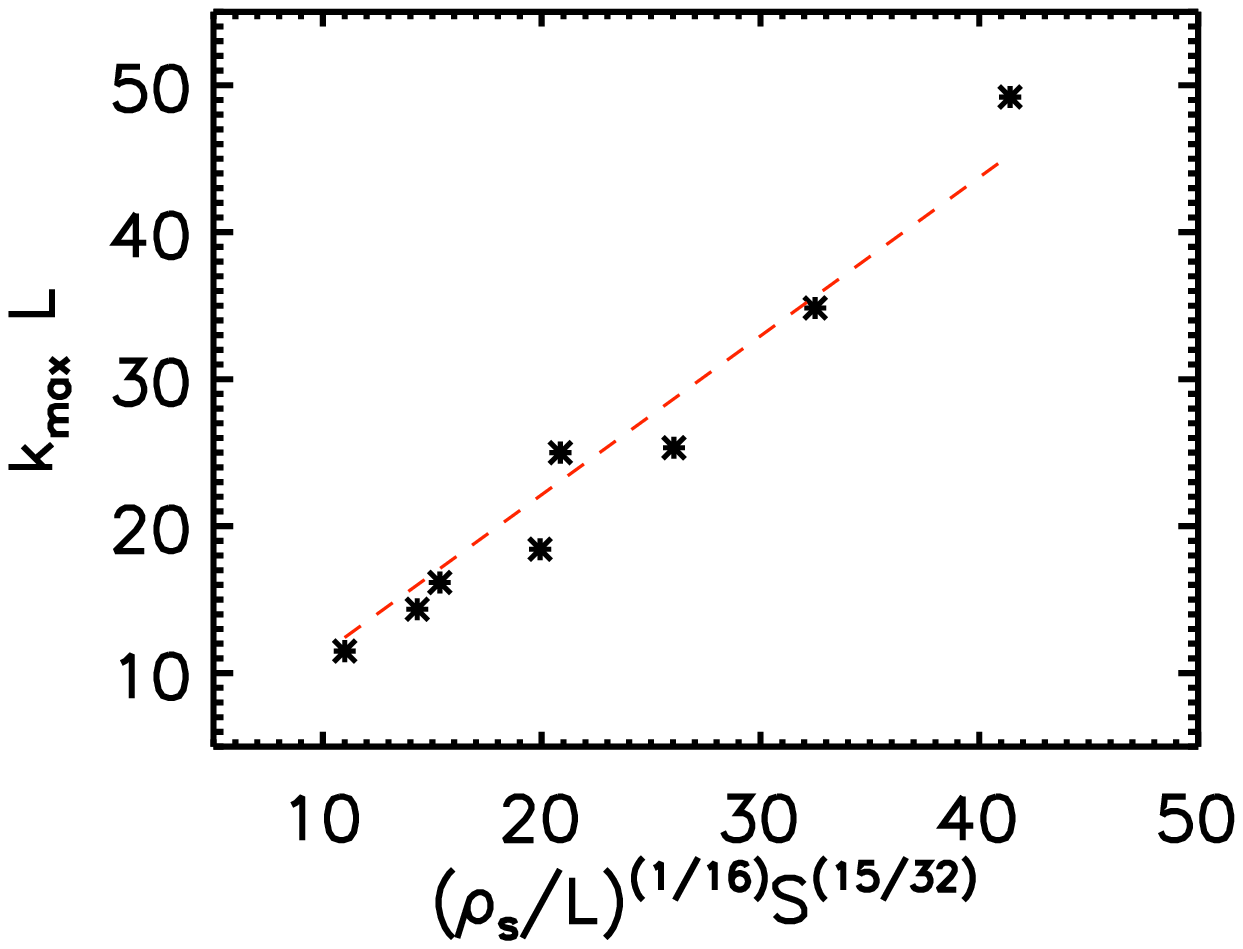}}
  \caption{Wavenumber measured from the width of the arising plasmoid in runs A-G, plotted against the theoretical prediction \Eq{wave2}}
  \label{kwid}
\end{figure}

It is noteworthy that all of the runs in this regime are at Lundquist numbers much lower than the MHD critical value of order $\sim 10^4$.  The lowest
$S$ at which a simulation (Run~A) obtains the plasmoid instability is ∼ 250.  As one decreases the value of $S$ even further we find in our
simulations that width of the primary island becomes as large as the simulation domain before any plasmoid is observed. In a less constrained system
it is possible that the plasmoid instability remains active at even lower values of $S$.

The linear regime of the plasmoid instability is very short-lived: it ends once the plasmoid becomes larger than the inner boundary layer. 
Although this layer is resolved in all our simulations, the number of grid points in the inner layer is not sufficiently 
large to be able to trace a long enough evolution in the linear regime. Thus, linear growth rate measurements cannot reliably be obtained. 
However, the wavenumber is a straightforward quantity to measure.
An intriguing observation is that only one single plasmoid arises in the runs A-H.  These runs range over a decade in Lundquist number and thus the
expected number of plasmoids varies by a factor of $\gtrsim 3$ from the run with lowest $S$ to the run with highest $S$, according to \Eq{wave2}.
That is not what we obtain, suggesting disagreement with the linear predictions.  This leads one to wonder why it is that the lower bound, $S >
(L/\rhos)^{14/9}$, is validated so well (Fig.~\ref{scbounds}), given that its validity relies on the scalings for $k_{max}$ (and $\gamma_{max}$) being
correct. 

To address this issue, we proceed as follows.  For each run, instead of counting the number of plasmoids to estimate the wavenumber, we compute it
from the measurement of the $y$-extent of the plasmoid (full width at half maximum of the island in $\Apar$) at the earliest possible stages of its appearance.  We find that
this measurement of the wavenumber does follow the theoretical scaling rather well, as shown in \Fig{kwid}.  We suspect that the explanation for this
result lies in the effects of spatial inhomogeneity in the direction along the sheet, as well as flows and reconnected component of the magnetic
field.  None of these effects is negligible at these low values of the Lundquist number, but they are all neglected in the theoretical derivation.
This argument is strengthened further by the results shown in appendix~\ref{validation}.  The important observation, however, is that the plasmoid
that does form has the correct wavenumber as predicted by linear theory.

\section{Discussions and conclusions}
In this paper, we have investigated, analytically and numerically, the semi-collisional regime of the plasmoid instability 
--- an extension to a Sweet-Parker sheet of the semi-collisional tearing mode~\citep{DL1977,CKH1986,zs2011}. 
We employ a reduced kinetic formalism obtainable from the gyrokinetic formalism via asymptotic expansion in low plasma $\beta$~\citep{zs2011}. 
The simulations are carried out with the code {\tt Viriato} \citep{nunoetal2016}.
This regime is analytically predicted to occupy a significant sliver of the reconnection phase diagram defined by a lower bound
$S> (L/\rhos)^{14/9}$ and an upper bound $S < (L/\rhos)^2$.
Our numerical simulations show that these bounds are remarkably robust:
runs which fall within these two bounds yield the plasmoid instability, and {\it vice versa}. 
The instability arises at much lower values of the Lundquist numbers than the MHD analog; as low as $S
\sim 250$.
We are limited in our exploration of even lower values of $S$ by our simulation setup. 
Thus, we do not rule out the formation of plasmoids for $S < 250$ in less constrained systems; indeed, we speculate 
that this regime could also potentially explain formation of plasmoids in recent reconnection experiments
\citep{jara-almonte_2016,hare2017prl,hare2017xiv,hare_experimental_17}. 
In Magpie, for example, we find from  Table I in \cite{HareEtal2018} that $S\approx 120$ and $L/d_i \approx 18$. If we were to consider the relevant kinetic scale to be $d_i$ instead 
of $\rhos$, then the value of these parameters ($S$ and $L/d_i$) satisfy the condition for the semi-collisional regime and fall within its bounds.
Note that in the collisionless (or weakly collisional) regime, plasmoids arise only when $L/\rhos$ is larger than 
a value of about $\sim 50$ (an empirical value seen in simulations) \citep{JD2011}, whereas the semi-collisional regime seems to offer no hard lower bound.
Further, with the validation of the existence of this regime, the reconnection phase diagram (Fig. 1 in \citet{JD2011}) would be modified. The lower bound of this regime, 
governed by either $S=(L/\rhos)^{14/9}$ or $S=(L/\rhos)^{8/5}$ would now cut across the two lines - one line representing the critical Lundquist number for MHD plasmoid instability ($S=S_c$)
and then a second line representing the lower bound of the multiple X-line hybrid regime ($S=L\sqrt{S_c}/2\rhos$). 

The numerical experiments reported here are limited in that the amount of flux to reconnect is finite, and the simulation box is periodic.
As such, (statistical) steady state reconnection cannot be attained, thereby preventing us from numerically answering the important question of what the
reconnection rate is in the semi-collisional plasmoid regime.
However, theoretically we may expect the following.
In the phase-space diagram of \Fig{scbounds}, 
assume that the initial system, with a certain $S$, $L$ and $\rho_s$, is in the semi-collisional regime.
As the plasmoid instability unfolds, we expect that smaller, inter-plasmoid current sheets will arise.  These will necessarily have a smaller length,
$L'\sim L/N$, where $N$ is the number of primary plasmoids.  Each of these inter-plasmoid current sheets now defines a reconnecting site which can be
located in the reconnection phase-diagram.  Since $\rho_s$ and $\eta$ are fixed, and assuming that $V_A$ is the same in these daughter sheets, the
only parameter that has changed is the length, from $L$ to $L'$.  This means a diagonal displacement in the direction of smaller $L/\rho_s$ from the
initial point in that diagram; the slope of that diagonal is $1$, because both axes are linearly proportional to $L$.  If the new position in this
diagram remains in the semi-collisional regime, each inter-plasmoid current sheet is still unstable to the semi-collisional plasmoid instability.  The
process then repeats (i.e., the plasmoid hierachy unfolds further) until arriving at an inter-plasmoid current sheet which is now short enough to be
outside of the semi-collisional bounds.  Inevitably, therefore, this lands the system to the left of the $S\sim (L/\rhos)^2$ bound, i.e., the
collisionless regime, where the expected reconnection rate is $\sim 0.1 \tau_A^{-1}$.
Defining $\lambda_c = (L_c/\rho_s)$ as an empirical scale separating the single from the multiple X-line collisionless regimes (numerically observed
to be $\sim 50$), we conclude that the system finally lands in either the  multiple, or single, X-line collisionless regime
depending on whether it is initially above or below the diagonal line $S\sim\lambda_c (L/\rhos)$, (which intersects the line $S \sim (L/\rhos)^2$ at
$(L/\rhos)=\lambda_c$) in the reconnection phase space diagram.

Finally, let us outline some general arguments for the case when, unlike our simulations, the global Lundquist number of the system is so large that a
Sweet-Parker current sheet may not be able to form dynamically.  Consider then a forming current sheet, and assume for simplicity that the
characteristic time at which it is forming is Alfv\'enic, $\tau_A = L/V_A$.  At any given moment of time, we parameterize the forming current sheet
aspect ratio as $a/L=S^{-\alpha}$~\cite{pucci_2014}, where $a$ is the current sheet width, and $\alpha$ is a number such that $0<\alpha<1/2$, with
$\alpha=1/2$ representing a Sweet-Parker sheet (which the system presumably never gets to).

Using \eq{gmax} (for a Harris-type sheet) we see that the growth rate of the  most unstable semi-collisional mode exceeds the current sheet formation
rate (Alfv\'enic) when
\be
\gamma_{max} \tau_A \sim (\rhos/L)^{2/3}S^{-1/3+2\alpha} \gtrsim 1.
\label{scgammaxn1}
\ee   
This expression constrains the relationship between $S$ and $L/\rhos$, for any value of $\alpha$, to attain Alfv\'enic growth at that value of $\alpha$. 
In addition, we must further require that this mode (the most unstable semi-collisional mode) is indeed in the semi-collisional regime, i.e., 
 $\delta^{SC}_{in, max} < \rhos$. This leads to, 
\be
(\rhos/L)^{1/9} S^{-(2\alpha/3) -2/9} < \rhos/L.
\label{scdelmaxn1}
\ee
These two expressions intersect when $\alpha=1/3$.
That is, if $\alpha<1/3$, the relationship that $S$ and $L/\rho_s$ must satisfy to yield Alfv\'enic growth is given by \eq{scgammaxn1}; if instead
$\alpha>1/3$, then it is sufficient to satisfy \eq{scdelmaxn1} to attain Alfv\'enic growth~\footnote{If instead one considers the case of a
sinusoidal-like current sheet, \eq{scgammaxn1} is replaced by $\gamma_{max} \tau_A \sim (\rhos/L)^{5/8}S^{-1/4+(15\alpha/8)} \gtrsim 1$; and 
\eq{scdelmaxn1} is replaced by $(\rhos/L)^{1/8} S^{-(5\alpha/8) -1/4} < \rhos/L$. Their intersection now takes place at $\alpha=3/10$. All the same arguments apply to this case.}.

A detailed analysis of all different possibilites that are encountered as $\alpha$ increases is beyond the scope of this paper and will be left to
future work.  Generally, for $\alpha<1/3$, the condition for the fastest semi-collisional mode to become faster than Alfv\'enic becomes progressively
less demanding on $S$.  Consider the particularly interesting case where $\alpha$ reaches the value of $1/3$. At this value, we must have
$S>(L/\rhos)^2$ for the semi-collisional mode to both exist and be super-Alfv\'enic.  Remarkably, $\alpha=1/3$ is the value at which the fastest
growing MHD mode becomes Alfv\'enic~\citep{pucci_2014}. The condition for that mode to indeed be in the MHD regime is, unsurprisingly, just the
reverse of the above, $S<(L/\rhos)^2$.  So, in this case ($\alpha=1/3$), the outcome is particularly simple: if $S>(L/\rho_s)^2$ the forming sheet would
be disrupted by a semi-collisional mode, and one might expect the reconnection rate to ensue to be $0.1\tau_A^{-1}$ as discussed above; if, instead
$S<(L/\rho_s)^2$ then the sheet would be disrupted by an MHD mode, and the reconnection rate would presumably be $S_c^{-1/2}\tau_A^{-1}\sim
0.01\tau_A^{-1}$ (unless a transition to the semi-collisional regime occurs during the plasmoid cascade -- see footnote on page \pageref{footnote}).
For example, in the solar corona, where $S\sim10^{13}$ and $L/\rho_s\sim 10^7$, we see that the MHD mode would win.

\acknowledgments
We thank Elizabeth Tolman and Ryan White for helpful discussions.  This work was supported by the NSF-DOE Partnership in Basic Plasma Science and
Engineering, Award No. DE-SC0016215.  The simulations presented in this paper were performed on the MIT-PSFC partition of the Engaging cluster at the
MGHPCC facility (www.mghpcc.org) which was funded by DoE grant number DE-FG02-91-ER54109.  We also acknowledge the usage of Stampede cluster in Texas
under the allocation TG-PHY140041.

\appendix \section{Semi-collisional tearing mode scalings in an SP sheet} \label{validation} 
\begin{figure}
  \center
  \includegraphics[width=0.45\textwidth]{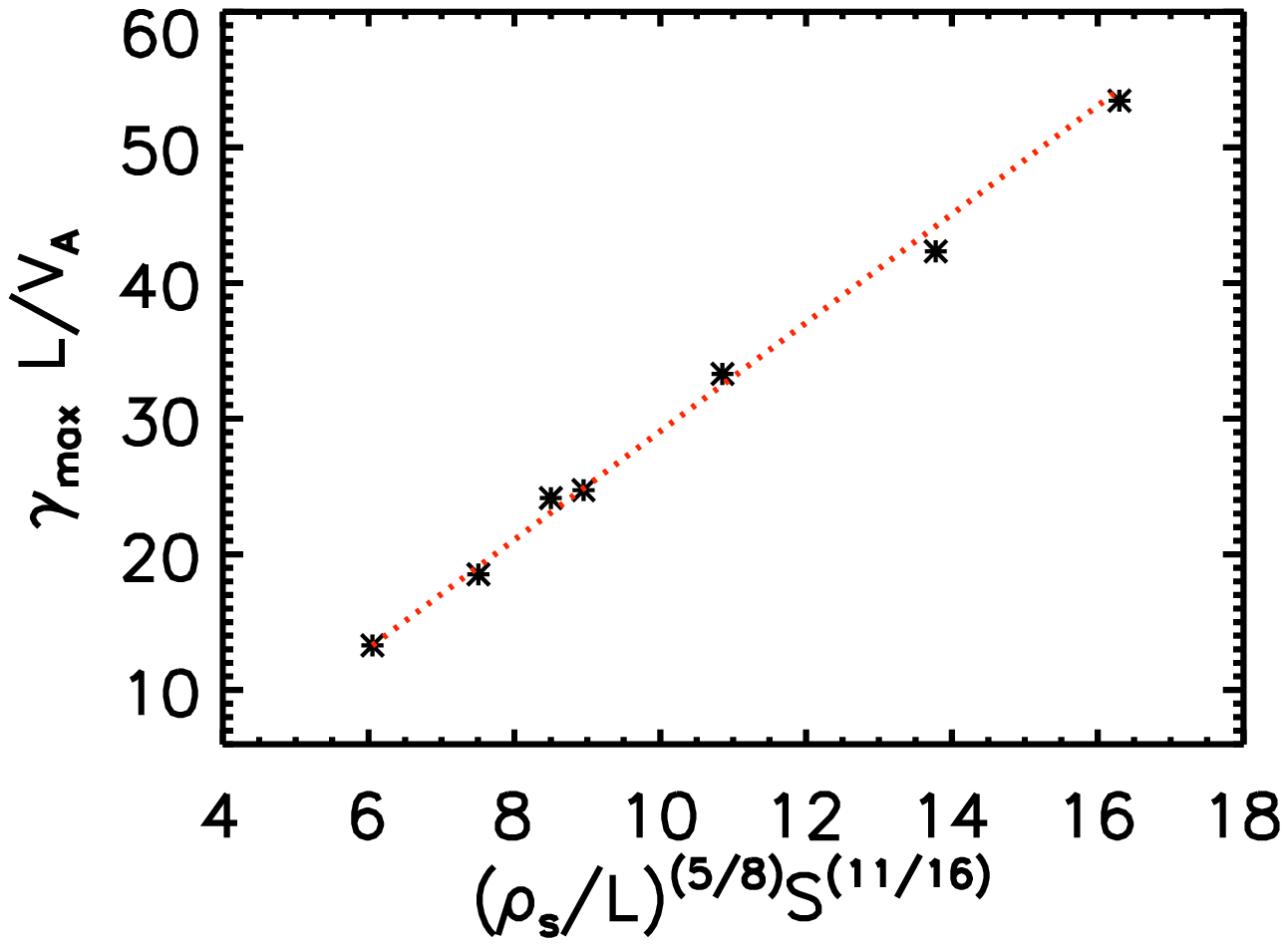}
  \includegraphics[width=0.45\textwidth]{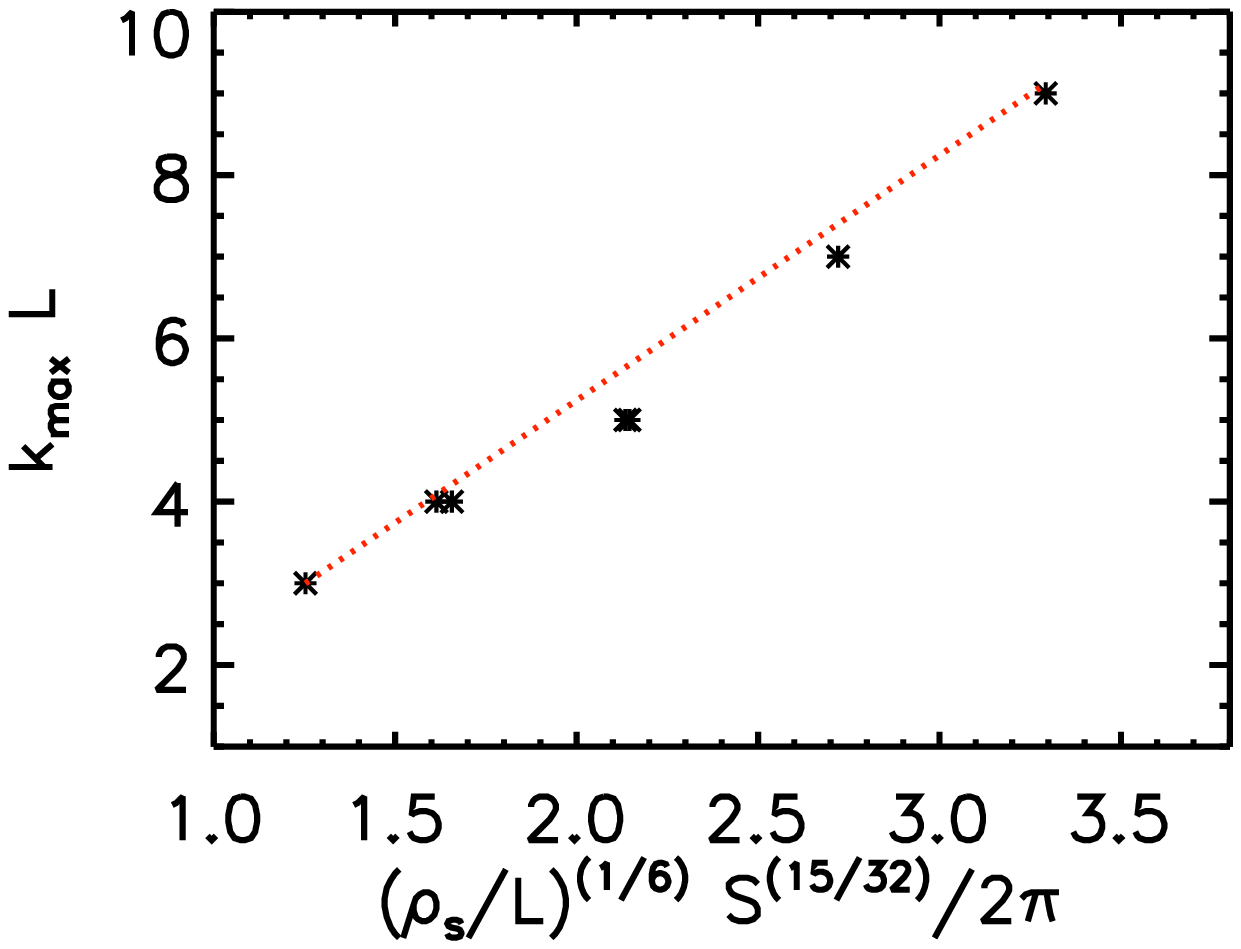}
  \caption{Growth rate (left) and wavenumber (right) of the most unstable semi-collisional plasmoid mode in a prescribed Sweet-Parker-like sheet, plotted against the theoretical predictions, \Eqs{gam2}{wave2}.}
  \label{SC_scalings}
\end{figure}

To confirm the validity of the analytical derivation of Section~\ref{sec:sc-plasmoid}, and the ability of the {\tt Viriato} code to recover the
scalings predicted there, we have performed a set of simulations whose key difference from those reported in the main text lies in the fact that the
thickness of initial magnetic profile is now the Sweet-Parker width, $L_y S^{-1/2}$, as is assumed throughout Section~\ref{sec:sc-plasmoid}.  This
initial configuration is not an exact Sweet-Parker sheet because it lacks both the appropriate flows, and the reconnected component of the magnetic
field; these are also simplifications adopted in the theoretical derivation, and which can be shown analytically to be
justifiable~\citep{nunoetal2007,loureiro_2013}.

All possible modes are seeded by the introduction of a small amplitude random
perturbation at $t=0$.  After an initial transient, we observe the exponential growth of a single mode --- the most unstable perturbation.
The growth rate and wavenumber (determined by counting the number of arising plasmoids) of this mode are plotted in Fig.~\ref{SC_scalings}. 
Excellent agreement with the theoretical scalings of  \Eqs{gam2}{wave2} is observed.

\bibliographystyle{jpp}
\bibliography{scpi}

\begin{thebibliography}{84}
\expandafter\ifx\csname natexlab\endcsname\relax\def\natexlab#1{#1}\fi
\def\au#1{#1} \def\ed#1{#1} \def\yr#1{#1}\def\at#1{#1}\def\jt#1{\textit{#1}}
  \def\bt#1{#1}\def\bvol#1{\textbf{#1}} \def\vol#1{#1} \def\pg#1{#1}
  \def\publ#1{#1}\def\arxiv#1{#1}\def\org#1{#1}\def\st#1{\textit{#1}}

\bibitem[{Baalrud} {\em et~al.\/}(2011){Baalrud}, {Bhattacharjee}, {Huang} \&
  {Germaschewski}]{baalrud_11}
{\sc \au{{Baalrud}, S.~D.}, \au{{Bhattacharjee}, A.}, \au{{Huang}, Y.-M.} \&
  \au{{Germaschewski}, K.}} \yr{2011}  \at{{Hall magnetohydrodynamic
  reconnection in the plasmoid unstable regime}}.  \jt{Physics of Plasmas}
  \bvol{18}~(9),  \pg{092108}.

\bibitem[{Baty}(2014)]{baty_2014}
{\sc \au{{Baty}, H.}} \yr{2014}  \at{{Effect of plasma-{$\beta$} on the onset
  of plasmoid instability in Sweet-Parker current sheets}}.  \jt{Journal of
  Plasma Physics}  \bvol{80},  \pg{655--665}.

\bibitem[{Bhattacharjee} {\em et~al.\/}(2009){Bhattacharjee}, {Huang}, {Yang}
  \& {Rogers}]{BHY2009}
{\sc \au{{Bhattacharjee}, A.}, \au{{Huang}, Y.-M.}, \au{{Yang}, H.} \&
  \au{{Rogers}, B.}} \yr{2009}  \at{{Fast reconnection in high-Lundquist-number
  plasmas due to the plasmoid instability}}.  \jt{Physics of Plasmas}
  \bvol{16}~(11),  \pg{112102}.

\bibitem[{Biskamp}(1986)]{biskamp86}
{\sc \au{{Biskamp}, D.}} \yr{1986}  \at{{Magnetic reconnection via current
  sheets}}.  \jt{Physics of Fluids}  \bvol{29},  \pg{1520--1531}.

\bibitem[{Biskamp}(2000)]{biskamp_book}
{\sc \au{{Biskamp}, D.}} \yr{2000} {\em {Magnetic Reconnection in Plasmas}\/}.
  \publ{Cambridge University Press}.

\bibitem[{Burch} {\em et~al.\/}(2016){Burch}, {Torbert}, {Phan}, {Chen},
  {Moore}, {Ergun}, {Eastwood}, {Gershman}, {Cassak}, {Argall}, {Wang},
  {Hesse}, {Pollock}, {Giles}, {Nakamura}, {Mauk}, {Fuselier}, {Russell},
  {Strangeway}, {Drake}, {Shay}, {Khotyaintsev}, {Lindqvist}, {Marklund},
  {Wilder}, {Young}, {Torkar}, {Goldstein}, {Dorelli}, {Avanov}, {Oka},
  {Baker}, {Jaynes}, {Goodrich}, {Cohen}, {Turner}, {Fennell}, {Blake},
  {Clemmons}, {Goldman}, {Newman}, {Petrinec}, {Trattner}, {Lavraud}, {Reiff},
  {Baumjohann}, {Magnes}, {Steller}, {Lewis}, {Saito}, {Coffey} \&
  {Chandler}]{burch_16}
{\sc \au{{Burch}, J.~L.}, \au{{Torbert}, R.~B.}, \au{{Phan}, T.~D.},
  \au{{Chen}, L.-J.}, \au{{Moore}, T.~E.}, \au{{Ergun}, R.~E.}, \au{{Eastwood},
  J.~P.}, \au{{Gershman}, D.~J.}, \au{{Cassak}, P.~A.}, \au{{Argall}, M.~R.},
  \au{{Wang}, S.}, \au{{Hesse}, M.}, \au{{Pollock}, C.~J.}, \au{{Giles},
  B.~L.}, \au{{Nakamura}, R.}, \au{{Mauk}, B.~H.}, \au{{Fuselier}, S.~A.},
  \au{{Russell}, C.~T.}, \au{{Strangeway}, R.~J.}, \au{{Drake}, J.~F.},
  \au{{Shay}, M.~A.}, \au{{Khotyaintsev}, Y.~V.}, \au{{Lindqvist}, P.-A.},
  \au{{Marklund}, G.}, \au{{Wilder}, F.~D.}, \au{{Young}, D.~T.}, \au{{Torkar},
  K.}, \au{{Goldstein}, J.}, \au{{Dorelli}, J.~C.}, \au{{Avanov}, L.~A.},
  \au{{Oka}, M.}, \au{{Baker}, D.~N.}, \au{{Jaynes}, A.~N.}, \au{{Goodrich},
  K.~A.}, \au{{Cohen}, I.~J.}, \au{{Turner}, D.~L.}, \au{{Fennell}, J.~F.},
  \au{{Blake}, J.~B.}, \au{{Clemmons}, J.}, \au{{Goldman}, M.}, \au{{Newman},
  D.}, \au{{Petrinec}, S.~M.}, \au{{Trattner}, K.~J.}, \au{{Lavraud}, B.},
  \au{{Reiff}, P.~H.}, \au{{Baumjohann}, W.}, \au{{Magnes}, W.}, \au{{Steller},
  M.}, \au{{Lewis}, W.}, \au{{Saito}, Y.}, \au{{Coffey}, V.} \& \au{{Chandler},
  M.}} \yr{2016}  \at{{Electron-scale measurements of magnetic reconnection in
  space}}.  \jt{Science}  \bvol{352},  \pg{aaf2939}.

\bibitem[{Cassak} {\em et~al.\/}(2009){Cassak}, {Shay} \& {Drake}]{cassak_2009}
{\sc \au{{Cassak}, P.~A.}, \au{{Shay}, M.~A.} \& \au{{Drake}, J.~F.}} \yr{2009}
   \at{{Scaling of Sweet-Parker reconnection with secondary islands}}.
  \jt{Physics of Plasmas}  \bvol{16}~(12),  \pg{120702}.

\bibitem[{Cerri} \& {Califano}(2017)]{cerri_17}
{\sc \au{{Cerri}, S.~S.} \& \au{{Califano}, F.}} \yr{2017}  \at{{Reconnection
  and small-scale fields in 2D-3V hybrid-kinetic driven turbulence
  simulations}}.  \jt{New Journal of Physics}  \bvol{19}~(2),  \pg{025007}.

\bibitem[{Cerutti} {\em et~al.\/}(2012){Cerutti}, {Uzdensky} \&
  {Begelman}]{ceruttietal12}
{\sc \au{{Cerutti}, B.}, \au{{Uzdensky}, D.~A.} \& \au{{Begelman}, M.~C.}}
  \yr{2012}  \at{{Extreme Particle Acceleration in Magnetic Reconnection
  Layers: Application to the Gamma-Ray Flares in the Crab Nebula}}.
  \jt{Astrophysical Journal}  \bvol{746},  \pg{148}.

\bibitem[{Cerutti} {\em et~al.\/}(2013){Cerutti}, {Werner}, {Uzdensky} \&
  {Begelman}]{ceruttietal2013}
{\sc \au{{Cerutti}, B.}, \au{{Werner}, G.~R.}, \au{{Uzdensky}, D.~A.} \&
  \au{{Begelman}, M.~C.}} \yr{2013}  \at{{Simulations of Particle Acceleration
  beyond the Classical Synchrotron Burnoff Limit in Magnetic Reconnection: An
  Explanation of the Crab Flares}}.  \jt{Astrophysical Journal}  \bvol{770},
  \pg{147}.

\bibitem[{Comisso} {\em et~al.\/}(2016){Comisso}, {Lingam}, {Huang} \&
  {Bhattacharjee}]{comisso_2016}
{\sc \au{{Comisso}, L.}, \au{{Lingam}, M.}, \au{{Huang}, Y.-M.} \&
  \au{{Bhattacharjee}, A.}} \yr{2016}  \at{{General theory of the plasmoid
  instability}}.  \jt{Physics of Plasmas}  \bvol{23}~(10),  \pg{100702}.

\bibitem[{Cowley} {\em et~al.\/}(1986){Cowley}, {Kulsrud} \& {Hahm}]{CKH1986}
{\sc \au{{Cowley}, S.~C.}, \au{{Kulsrud}, R.~M.} \& \au{{Hahm}, T.~S.}}
  \yr{1986}  \at{{Linear stability of tearing modes}}.  \jt{Physics of Fluids}
  \bvol{29},  \pg{3230--3244}.

\bibitem[{Daughton} \& {Roytershteyn}(2012)]{DR2012}
{\sc \au{{Daughton}, W.} \& \au{{Roytershteyn}, V.}} \yr{2012}  \at{{Emerging
  Parameter Space Map of Magnetic Reconnection in Collisional and Kinetic
  Regimes}}.  \jt{\ssr}  \bvol{172},  \pg{271--282}.

\bibitem[{DiBraccio} {\em et~al.\/}(2015){DiBraccio}, {Slavin}, {Imber},
  {Gershman}, {Raines}, {Jackman}, {Boardsen}, {Anderson}, {Korth},
  {Zurbuchen}, {McNutt} \& {Solomon}]{Braccioetal15}
{\sc \au{{DiBraccio}, G.~A.}, \au{{Slavin}, J.~A.}, \au{{Imber}, S.~M.},
  \au{{Gershman}, D.~J.}, \au{{Raines}, J.~M.}, \au{{Jackman}, C.~M.},
  \au{{Boardsen}, S.~A.}, \au{{Anderson}, B.~J.}, \au{{Korth}, H.},
  \au{{Zurbuchen}, T.~H.}, \au{{McNutt}, R.~L.} \& \au{{Solomon}, S.~C.}}
  \yr{2015}  \at{{MESSENGER observations of flux ropes in Mercury's
  magnetotail}}.  \jt{Planetary and Space Science}  \bvol{115},  \pg{77--89}.

\bibitem[{Dorfman} {\em et~al.\/}(2013){Dorfman}, {Ji}, {Yamada}, {Yoo},
  {Lawrence}, {Myers} \& {Tharp}]{dorfman_2013}
{\sc \au{{Dorfman}, S.}, \au{{Ji}, H.}, \au{{Yamada}, M.}, \au{{Yoo}, J.},
  \au{{Lawrence}, E.}, \au{{Myers}, C.} \& \au{{Tharp}, T.~D.}} \yr{2013}
  \at{{Three-dimensional, impulsive magnetic reconnection in a laboratory
  plasma}}.  \jt{Geophysical Review Letters}  \bvol{40},  \pg{233--238}.

\bibitem[{Drake} \& {Lee}(1977)]{DL1977}
{\sc \au{{Drake}, J.~F.} \& \au{{Lee}, Y.~C.}} \yr{1977}  \at{{Kinetic theory
  of tearing instabilities}}.  \jt{Physics of Fluids}  \bvol{20},
  \pg{1341--1353}.

\bibitem[{Drake} {\em et~al.\/}(2006){Drake}, {Swisdak}, {Che} \&
  {Shay}]{drake_2006}
{\sc \au{{Drake}, J.~F.}, \au{{Swisdak}, M.}, \au{{Che}, H.} \& \au{{Shay},
  M.~A.}} \yr{2006}  \at{{Electron acceleration from contracting magnetic
  islands during reconnection}}.  \jt{Nature}  \bvol{443},  \pg{553--556}.

\bibitem[{Dungey}(1961)]{dungey_1961}
{\sc \au{{Dungey}, J.~W.}} \yr{1961}  \at{{Interplanetary Magnetic Field and
  the Auroral Zones}}.  \jt{Physical Review Letters}  \bvol{6},  \pg{47--48}.

\bibitem[{Egedal} {\em et~al.\/}(2014){Egedal}, {Olson}, {Endrizzi} \&
  {Forest}]{trex}
{\sc \au{{Egedal}, J.}, \au{{Olson}, J.}, \au{{Endrizzi}, D.} \& \au{{Forest},
  C.}} \yr{2014}  \at{{Opportunities in TREX, a New Terrestrial Reconnection
  EXperiment.}}  \jt{AGU Fall Meeting Abstracts} .

\bibitem[{Fox} {\em et~al.\/}(2012){Fox}, {Bhattacharjee} \&
  {Germaschewski}]{foxetal2011}
{\sc \au{{Fox}, W.}, \au{{Bhattacharjee}, A.} \& \au{{Germaschewski}, K.}}
  \yr{2012}  \at{{Magnetic reconnection in high-energy-density laser-produced
  plasmasa)}}.  \jt{Physics of Plasmas}  \bvol{19}~(5),  \pg{056309}.

\bibitem[{Frieman} \& {Chen}(1982)]{FC1982}
{\sc \au{{Frieman}, E.~A.} \& \au{{Chen}, L.}} \yr{1982}  \at{{Nonlinear
  gyrokinetic equations for low-frequency electromagnetic waves in general
  plasma equilibria}}.  \jt{Physics of Fluids}  \bvol{25},  \pg{502--508}.

\bibitem[{Furth} {\em et~al.\/}(1963){Furth}, {Killeen} \& {Rosenbluth}]{FKR}
{\sc \au{{Furth}, H.~P.}, \au{{Killeen}, J.} \& \au{{Rosenbluth}, M.~N.}}
  \yr{1963}  \at{{Finite-Resistivity Instabilities of a Sheet Pinch}}.
  \jt{Physics of Fluids}  \bvol{6},  \pg{459--484}.

\bibitem[{Giannios} {\em et~al.\/}(2009){Giannios}, {Uzdensky} \&
  {Begelman}]{gianniaosetal2009}
{\sc \au{{Giannios}, D.}, \au{{Uzdensky}, D.~A.} \& \au{{Begelman}, M.~C.}}
  \yr{2009}  \at{{Fast TeV variability in blazars: jets in a jet}}.
  \jt{Monthly Notices of the Royal Astronomical Society}  \bvol{395},
  \pg{L29--L33}.

\bibitem[{Guo} {\em et~al.\/}(2015){Guo}, {Liu}, {Daughton} \&
  {Li}]{guoetal2015}
{\sc \au{{Guo}, F.}, \au{{Liu}, Y.-H.}, \au{{Daughton}, W.} \& \au{{Li}, H.}}
  \yr{2015}  \at{{Particle Acceleration and Plasma Dynamics during Magnetic
  Reconnection in the Magnetically Dominated Regime}}.  \jt{Astrophysical
  Journal}  \bvol{806},  \pg{167}.

\bibitem[{Hare}(2017)]{hare_thesis}
{\sc \au{{Hare}, J.~D.}} \yr{2017}  \at{High energy density magnetic
  reconnection experiments in colliding carbon plasma flows}. PhD thesis,
  Imperial College London.

\bibitem[{Hare} {\em et~al.\/}(2017{\natexlab{{\em a\/}}}){Hare},
  {\GG{20020530}}{Suttle}, {Lebedev}, {Loureiro}, {Ciardi}, {Burdiak},
  {Chittenden}, {Clayson}, {Garcia}, {Niasse}, {Robinson}, {Smith}, {Stuart},
  {Suzuki-Vidal}, {Swadling}, {Ma}, {Wu} \& {Yang}]{hare2017prl}
{\sc \au{{Hare}, J.~D.}, \au{{\GG{20020530}}{Suttle}, L.}, \au{{Lebedev},
  S.~V.}, \au{{Loureiro}, N.~F.}, \au{{Ciardi}, A.}, \au{{Burdiak}, G.~C.},
  \au{{Chittenden}, J.~P.}, \au{{Clayson}, T.}, \au{{Garcia}, C.},
  \au{{Niasse}, N.}, \au{{Robinson}, T.}, \au{{Smith}, R.~A.}, \au{{Stuart},
  N.}, \au{{Suzuki-Vidal}, F.}, \au{{Swadling}, G.~F.}, \au{{Ma}, J.},
  \au{{Wu}, J.} \& \au{{Yang}, Q.}} \yr{2017{\natexlab{{\em a\/}}}}
  \at{{Anomalous Heating and Plasmoid Formation in a Driven Magnetic
  Reconnection Experiment}}.  \jt{Physical Review Letters}  \bvol{118}~(8),
  \pg{085001}.

\bibitem[{Hare} {\em et~al.\/}(2017{\natexlab{{\em b\/}}}){Hare},
  {\GG{20020603}}{Lebedev}, {Suttle}, {Loureiro}, {Ciardi}, {Burdiak},
  {Chittenden}, {Clayson}, {Eardley}, {Garcia}, {Halliday}, {Niasse},
  {Robinson}, {Smith}, {Stuart}, {Suzuki-Vidal}, {Swadling}, {Ma} \&
  {Wu}]{hare2017xiv}
{\sc \au{{Hare}, J.~D.}, \au{{\GG{20020603}}{Lebedev}, S.~V.}, \au{{Suttle},
  L.~G.}, \au{{Loureiro}, N.~F.}, \au{{Ciardi}, A.}, \au{{Burdiak}, G.~C.},
  \au{{Chittenden}, J.~P.}, \au{{Clayson}, T.}, \au{{Eardley}, S.~J.},
  \au{{Garcia}, C.}, \au{{Halliday}, J.~W.~D.}, \au{{Niasse}, N.},
  \au{{Robinson}, T.}, \au{{Smith}, R.~A.}, \au{{Stuart}, N.},
  \au{{Suzuki-Vidal}, F.}, \au{{Swadling}, G.~F.}, \au{{Ma}, J.} \& \au{{Wu},
  J.}} \yr{2017{\natexlab{{\em b\/}}}}  \at{{Formation and structure of a
  current sheet in pulsed-power driven magnetic reconnection experiments}}.
  \jt{Physics of Plasmas}  \bvol{24}~(10),  \pg{102703}.

\bibitem[{Hare} {\em et~al.\/}(2018{\natexlab{{\em a\/}}}){Hare}, {Suttle},
  {Lebedev}, {Loureiro}, {Ciardi}, {Chittenden}, {Clayson}, {Eardley},
  {Garcia}, {Halliday}, {Robinson}, {Smith}, {Stuart}, {Suzuki-Vidal} \&
  {Tubman}]{hare_experimental_17}
{\sc \au{{Hare}, J.~D.}, \au{{Suttle}, L.~G.}, \au{{Lebedev}, S.~V.},
  \au{{Loureiro}, N.~F.}, \au{{Ciardi}, A.}, \au{{Chittenden}, J.~P.},
  \au{{Clayson}, T.}, \au{{Eardley}, S.~J.}, \au{{Garcia}, C.}, \au{{Halliday},
  J.~W.~D.}, \au{{Robinson}, T.}, \au{{Smith}, R.~A.}, \au{{Stuart}, N.},
  \au{{Suzuki-Vidal}, F.} \& \au{{Tubman}, E.~R.}} \yr{2018{\natexlab{{\em
  a\/}}}}  \at{{An experimental platform for pulsed-power driven magnetic
  reconnection}}.  \jt{Physics of Plasmas}  \bvol{25}~(5),  \pg{055703}.

\bibitem[{Hare} {\em et~al.\/}(2018{\natexlab{{\em b\/}}}){Hare}, {Suttle},
  {Lebedev}, {Loureiro}, {Ciardi}, {Chittenden}, {Clayson}, {Eardley},
  {Garcia}, {Halliday}, {Robinson}, {Smith}, {Stuart}, {Suzuki-Vidal} \&
  {Tubman}]{HareEtal2018}
{\sc \au{{Hare}, J.~D.}, \au{{Suttle}, L.~G.}, \au{{Lebedev}, S.~V.},
  \au{{Loureiro}, N.~F.}, \au{{Ciardi}, A.}, \au{{Chittenden}, J.~P.},
  \au{{Clayson}, T.}, \au{{Eardley}, S.~J.}, \au{{Garcia}, C.}, \au{{Halliday},
  J.~W.~D.}, \au{{Robinson}, T.}, \au{{Smith}, R.~A.}, \au{{Stuart}, N.},
  \au{{Suzuki-Vidal}, F.} \& \au{{Tubman}, E.~R.}} \yr{2018{\natexlab{{\em
  b\/}}}}  \at{{An experimental platform for pulsed-power driven magnetic
  reconnection}}.  \jt{Physics of Plasmas}  \bvol{25}~(5),  \pg{055703}.

\bibitem[{Harris}(1962)]{harris62}
{\sc \au{{Harris}, E.~G.}} \yr{1962}  \at{{On a plasma sheath separating
  regions of oppositely directed magnetic field}}.  \jt{Il Nuovo Cimento}
  \bvol{23},  \pg{115--121}.

\bibitem[{Hastie}(1997)]{hastie_sawtooth_1997}
{\sc \au{{Hastie}, R.~J.}} \yr{1997}  \at{{Sawtooth instability in tokamak
  plasmas}}.  \jt{Astrophysics and Space Science}  \bvol{256},  \pg{177--204}.

\bibitem[{Howes} {\em et~al.\/}(2006){Howes}, {Cowley}, {Dorland}, {Hammett},
  {Quataert} \& {Schekochihin}]{howesetal2006}
{\sc \au{{Howes}, G.~G.}, \au{{Cowley}, S.~C.}, \au{{Dorland}, W.},
  \au{{Hammett}, G.~W.}, \au{{Quataert}, E.} \& \au{{Schekochihin}, A.~A.}}
  \yr{2006}  \at{{Astrophysical Gyrokinetics: Basic Equations and Linear
  Theory}}.  \jt{Astrophysical Journal}  \bvol{651},  \pg{590--614}.

\bibitem[{Huang} \& {Bhattacharjee}(2010)]{huang_2010}
{\sc \au{{Huang}, Y.-M.} \& \au{{Bhattacharjee}, A.}} \yr{2010}  \at{{Scaling
  laws of resistive magnetohydrodynamic reconnection in the
  high-Lundquist-number, plasmoid-unstable regime}}.  \jt{Physics of Plasmas}
  \bvol{17}~(6),  \pg{062104--062104}.

\bibitem[{Jackman} {\em et~al.\/}(2011){Jackman}, {Slavin} \&
  {Cowley}]{jackmanetal2011}
{\sc \au{{Jackman}, C.~M.}, \au{{Slavin}, J.~A.} \& \au{{Cowley}, S.~W.~H.}}
  \yr{2011}  \at{{Cassini observations of plasmoid structure and dynamics:
  Implications for the role of magnetic reconnection in magnetospheric
  circulation at Saturn}}.  \jt{Journal of Geophysical Research (Space
  Physics)}  \bvol{116},  \pg{A10212}.

\bibitem[{Jara-Almonte} {\em et~al.\/}(2016){Jara-Almonte}, {Ji}, {Yamada},
  {Yoo} \& {Fox}]{jara-almonte_2016}
{\sc \au{{Jara-Almonte}, J.}, \au{{Ji}, H.}, \au{{Yamada}, M.}, \au{{Yoo}, J.}
  \& \au{{Fox}, W.}} \yr{2016}  \at{{Laboratory Observation of Resistive
  Electron Tearing in a Two-Fluid Reconnecting Current Sheet}}.  \jt{Physical
  Review Letters}  \bvol{117}~(9),  \pg{095001}.

\bibitem[{Ji} {\em et~al.\/}(2015){Ji}, {Bhattacharjee}, {Prager}, {Daughton},
  {Bale}, {Carter}, {Crocker}, {Drake}, {Egedal}, {Sarff}, {Fox},
  {Jara-Almonte}, {Myers}, {Ren}, {Yamada} \& {Yoo}]{flare}
{\sc \au{{Ji}, H.}, \au{{Bhattacharjee}, A.}, \au{{Prager}, S.},
  \au{{Daughton}, W.}, \au{{Bale}, S.~D.}, \au{{Carter}, T.}, \au{{Crocker},
  N.}, \au{{Drake}, J.}, \au{{Egedal}, J.}, \au{{Sarff}, J.}, \au{{Fox}, W.},
  \au{{Jara-Almonte}, J.}, \au{{Myers}, C.}, \au{{Ren}, Y.}, \au{{Yamada}, M.}
  \& \au{{Yoo}, J.}} \yr{2015} {FLARE (Facility for Laboratory Reconnection
  Experiments): A Major Next-Step for Laboratory Studies of Magnetic
  Reconnection}.  \bt{In {\em AAS/AGU Triennial Earth-Sun Summit\/}},
  \st{AAS/AGU Triennial Earth-Sun Summit},  \vol{vol.~1},  \pg{p. 104.05}.

\bibitem[{Ji} \& {Daughton}(2011)]{JD2011}
{\sc \au{{Ji}, H.} \& \au{{Daughton}, W.}} \yr{2011}  \at{{Phase diagram for
  magnetic reconnection in heliophysical, astrophysical, and laboratory
  plasmas}}.  \jt{Physics of Plasmas}  \bvol{18}~(11),  \pg{111207--111207}.

\bibitem[Kagan {\em et~al.\/}(2015)Kagan, Sironi, Cerutti \&
  Giannios]{Kagan2015}
{\sc \au{Kagan, D.}, \au{Sironi, L.}, \au{Cerutti, B.} \& \au{Giannios, D.}}
  \yr{2015}  \at{Relativistic magnetic reconnection in pair plasmas and its
  astrophysical applications}.  \jt{Space Science Reviews}  \bvol{191}~(1),
  \pg{545--573}.

\bibitem[{Lapenta}(2008)]{lapenta_2008}
{\sc \au{{Lapenta}, G.}} \yr{2008}  \at{{Self-Feeding Turbulent Magnetic
  Reconnection on Macroscopic Scales}}.  \jt{Physical Review Letters}
  \bvol{100}~(23),  \pg{235001}.

\bibitem[{Lenard} \& {Bernstein}(1958)]{LenBern}
{\sc \au{{Lenard}, A.} \& \au{{Bernstein}, I.~B.}} \yr{1958}  \at{{Plasma
  Oscillations with Diffusion in Velocity Space}}.  \jt{Physical Review}
  \bvol{112},  \pg{1456--1459}.

\bibitem[{Liu} {\em et~al.\/}(2013){Liu}, {Chen} \& {Petrosian}]{liuetal13}
{\sc \au{{Liu}, W.}, \au{{Chen}, Q.} \& \au{{Petrosian}, V.}} \yr{2013}
  \at{{Plasmoid Ejections and Loop Contractions in an Eruptive M7.7 Solar
  Flare: Evidence of Particle Acceleration and Heating in Magnetic Reconnection
  Outflows}}.  \jt{Astrophysical Journal}  \bvol{767},  \pg{168}.

\bibitem[{Loureiro} \& {Boldyrev}(2017)]{loureiro_role_2017}
{\sc \au{{Loureiro}, N.~F.} \& \au{{Boldyrev}, S.}} \yr{2017}
  \at{{{\GG{20020530}}Role of Magnetic Reconnection in Magnetohydrodynamic
  Turbulence}}.  \jt{Physical Review Letters}  \bvol{118}~(24),  \pg{245101}.

\bibitem[{Loureiro} {\em et~al.\/}(2005){Loureiro}, {Cowley}, {Dorland},
  {Haines} \& {Schekochihin}]{nunoetal2005}
{\sc \au{{Loureiro}, N.~F.}, \au{{Cowley}, S.~C.}, \au{{Dorland}, W.~D.},
  \au{{Haines}, M.~G.} \& \au{{Schekochihin}, A.~A.}} \yr{2005}  \at{{X-Point
  Collapse and Saturation in the Nonlinear Tearing Mode Reconnection}}.
  \jt{Physical Review Letters}  \bvol{95}~(23),  \pg{235003}.

\bibitem[{Loureiro} {\em et~al.\/}(2016){Loureiro}, {Dorland}, {Fazendeiro},
  {Kanekar}, {Mallet}, {Vilelas} \& {Zocco}]{nunoetal2016}
{\sc \au{{Loureiro}, N.~F.}, \au{{Dorland}, W.}, \au{{Fazendeiro}, L.},
  \au{{Kanekar}, A.}, \au{{Mallet}, A.}, \au{{Vilelas}, M.~S.} \& \au{{Zocco},
  A.}} \yr{2016}  \at{{Viriato: A Fourier-Hermite spectral code for strongly
  magnetized fluid-kinetic plasma dynamics}}.  \jt{Computer Physics
  Communications}  \bvol{206},  \pg{45--63}.

\bibitem[{Loureiro} {\em et~al.\/}(2012){Loureiro}, {Samtaney}, {Schekochihin}
  \& {Uzdensky}]{loureiro_2012}
{\sc \au{{Loureiro}, N.~F.}, \au{{Samtaney}, R.}, \au{{Schekochihin}, A.~A.} \&
  \au{{Uzdensky}, D.~A.}} \yr{2012}  \at{{Magnetic reconnection and stochastic
  plasmoid chains in high-Lundquist-number plasmas}}.  \jt{Physics of Plasmas}
  \bvol{19}~(4),  \pg{042303--042303}.

\bibitem[{Loureiro} {\em et~al.\/}(2007){Loureiro}, {Schekochihin} \&
  {Cowley}]{nunoetal2007}
{\sc \au{{Loureiro}, N.~F.}, \au{{Schekochihin}, A.~A.} \& \au{{Cowley},
  S.~C.}} \yr{2007}  \at{{Instability of current sheets and formation of
  plasmoid chains}}.  \jt{Physics of Plasmas}  \bvol{14}~(10),
  \pg{100703--100703}.

\bibitem[{Loureiro} {\em et~al.\/}(2013){Loureiro}, {Schekochihin} \&
  {Uzdensky}]{loureiro_2013}
{\sc \au{{Loureiro}, N.~F.}, \au{{Schekochihin}, A.~A.} \& \au{{Uzdensky},
  D.~A.}} \yr{2013}  \at{{Plasmoid and Kelvin-Helmholtz instabilities in
  Sweet-Parker current sheets}}.  \jt{Physical Review E}  \bvol{87}~(1),
  \pg{013102}.

\bibitem[{Loureiro} \& {Uzdensky}(2016)]{nunouzdensky2016}
{\sc \au{{Loureiro}, N.~F.} \& \au{{Uzdensky}, D.~A.}} \yr{2016}  \at{{Magnetic
  reconnection: from the Sweet-Parker model to stochastic plasmoid chains}}.
  \jt{Plasma Physics and Controlled Fusion}  \bvol{58}~(1),  \pg{014021}.

\bibitem[{Mallet} {\em et~al.\/}(2017){Mallet}, {Schekochihin} \&
  {Chandran}]{mallet2017}
{\sc \au{{Mallet}, A.}, \au{{Schekochihin}, A.~A.} \& \au{{Chandran},
  B.~D.~G.}} \yr{2017}  \at{{{\GG{20020530}}Disruption of sheet-like structures
  in Alfvenic turbulence by magnetic reconnection}}.  \jt{Monthly Notices of
  the Royal Astronomical Society}  \bvol{468},  \pg{4862--4871}.

\bibitem[{Matthaeus} \& {Lamkin}(1986)]{matthaeus_86}
{\sc \au{{Matthaeus}, W.~H.} \& \au{{Lamkin}, S.~L.}} \yr{1986}  \at{{Turbulent
  magnetic reconnection}}.  \jt{Physics of Fluids}  \bvol{29},
  \pg{2513--2534}.

\bibitem[{Milligan} {\em et~al.\/}(2010){Milligan}, {McAteer}, {Dennis} \&
  {Young}]{milliganetal10}
{\sc \au{{Milligan}, R.~O.}, \au{{McAteer}, R.~T.~J.}, \au{{Dennis}, B.~R.} \&
  \au{{Young}, C.~A.}} \yr{2010}  \at{{Evidence of a Plasmoid-Looptop
  Interaction and Magnetic Inflows During a Solar Flare/Coronal Mass Ejection
  Eruptive Event}}.  \jt{Astrophysical Journal}  \bvol{713},  \pg{1292--1300}.

\bibitem[{Moldwin} \& {Hughes}(1992)]{moldwinHughes92}
{\sc \au{{Moldwin}, M.~B.} \& \au{{Hughes}, W.~J.}} \yr{1992}  \at{{On the
  formation and evolution of plasmoids - A survey of ISEE 3 Geotail data}}.
  \jt{\jgr}  \bvol{97},  \pg{19}.

\bibitem[{Nishizuka} {\em et~al.\/}(2010){Nishizuka}, {Takasaki}, {Asai} \&
  {Shibata}]{nishizukaetal2010}
{\sc \au{{Nishizuka}, N.}, \au{{Takasaki}, H.}, \au{{Asai}, A.} \&
  \au{{Shibata}, K.}} \yr{2010}  \at{{Multiple Plasmoid Ejections and
  Associated Hard X-ray Bursts in the 2000 November 24 Flare}}.
  \jt{Astrophysical Journal}  \bvol{711},  \pg{1062--1072}.

\bibitem[{Numata} \& {Loureiro}(2015)]{numata_2015}
{\sc \au{{Numata}, R.} \& \au{{Loureiro}, N.~F.}} \yr{2015}  \at{{Ion and
  electron heating during magnetic reconnection in weakly collisional
  plasmas}}.  \jt{Journal of Plasma Physics}  \bvol{81}~(2),  \pg{305810201}.

\bibitem[Oka {\em et~al.\/}(2010)Oka, Phan, Krucker, Fujimoto \&
  Shinohara]{Okaetal2010}
{\sc \au{Oka, M.}, \au{Phan, T.-D.}, \au{Krucker, S.}, \au{Fujimoto, M.} \&
  \au{Shinohara, I.}} \yr{2010}  \at{Electron acceleration by multi-island
  coalescence}.  \jt{The Astrophysical Journal}  \bvol{714}~(1),  \pg{915}.

\bibitem[{Olson} {\em et~al.\/}(2016{\natexlab{{\em a\/}}}){Olson}, {Egedal},
  {Greess}, {Myers}, {Clark}, {Endrizzi}, {Flanagan}, {Milhone}, {Peterson},
  {Wallace}, {Weisberg} \& {Forest}]{olson_2016}
{\sc \au{{Olson}, J.}, \au{{Egedal}, J.}, \au{{Greess}, S.}, \au{{Myers}, R.},
  \au{{Clark}, M.}, \au{{Endrizzi}, D.}, \au{{Flanagan}, K.}, \au{{Milhone},
  J.}, \au{{Peterson}, E.}, \au{{Wallace}, J.}, \au{{Weisberg}, D.} \&
  \au{{Forest}, C.~B.}} \yr{2016{\natexlab{{\em a\/}}}}  \at{{Experimental
  Demonstration of the Collisionless Plasmoid Instability below the Ion Kinetic
  Scale during Magnetic Reconnection}}.  \jt{Physical Review Letters}
  \bvol{116}~(25),  \pg{255001}.

\bibitem[{Olson} {\em et~al.\/}(2016{\natexlab{{\em b\/}}}){Olson}, {Egedal},
  {Greess}, {Myers}, {Clark}, {Endrizzi}, {Flanagan}, {Milhone}, {Peterson},
  {Wallace}, {Weisberg} \& {Forest}]{olsonetal2016}
{\sc \au{{Olson}, J.}, \au{{Egedal}, J.}, \au{{Greess}, S.}, \au{{Myers}, R.},
  \au{{Clark}, M.}, \au{{Endrizzi}, D.}, \au{{Flanagan}, K.}, \au{{Milhone},
  J.}, \au{{Peterson}, E.}, \au{{Wallace}, J.}, \au{{Weisberg}, D.} \&
  \au{{Forest}, C.~B.}} \yr{2016{\natexlab{{\em b\/}}}}  \at{{Experimental
  Demonstration of the Collisionless Plasmoid Instability below the Ion Kinetic
  Scale during Magnetic Reconnection}}.  \jt{Physical Review Letters}
  \bvol{116}~(25),  \pg{255001}.

\bibitem[{Parker}(1957)]{parker_1957}
{\sc \au{{Parker}, E.~N.}} \yr{1957}  \at{{Sweet's Mechanism for Merging
  Magnetic Fields in Conducting Fluids}}.  \jt{Journal of Geophysical Research}
   \bvol{62},  \pg{509--520}.

\bibitem[{Pegoraro} \& {Schep}(1986)]{pegoraro1986}
{\sc \au{{Pegoraro}, F.} \& \au{{Schep}, T.~J.}} \yr{1986}  \at{{Theory of
  resistive modes in the ballooning representation}}.  \jt{Plasma Physics and
  Controlled Fusion}  \bvol{28},  \pg{647--667}.

\bibitem[{Pucci} \& {Velli}(2014)]{pucci_2014}
{\sc \au{{Pucci}, F.} \& \au{{Velli}, M.}} \yr{2014}  \at{{Reconnection of
  Quasi-singular Current Sheets: The ``Ideal'' Tearing Mode}}.  \jt{The
  Astrophysical Journal Letters}  \bvol{780},  \pg{L19}.

\bibitem[{Samtaney} {\em et~al.\/}(2009){Samtaney}, {Loureiro}, {Uzdensky},
  {Schekochihin} \& {Cowley}]{samtaney_2009}
{\sc \au{{Samtaney}, R.}, \au{{Loureiro}, N.~F.}, \au{{Uzdensky}, D.~A.},
  \au{{Schekochihin}, A.~A.} \& \au{{Cowley}, S.~C.}} \yr{2009}  \at{{Formation
  of Plasmoid Chains in Magnetic Reconnection}}.  \jt{Physical Review Letters}
  \bvol{103}~(10),  \pg{105004}.

\bibitem[{Servidio} {\em et~al.\/}(2009){Servidio}, {Matthaeus}, {Shay},
  {Cassak} \& {Dmitruk}]{servidio_09}
{\sc \au{{Servidio}, S.}, \au{{Matthaeus}, W.~H.}, \au{{Shay}, M.~A.},
  \au{{Cassak}, P.~A.} \& \au{{Dmitruk}, P.}} \yr{2009}  \at{{Magnetic
  Reconnection in Two-Dimensional Magnetohydrodynamic Turbulence}}.
  \jt{Physical Review Letters}  \bvol{102}~(11),  \pg{115003}.

\bibitem[{Sharma} {\em et~al.\/}(2017){Sharma}, {Mitra} \&
  {Oberoi}]{sharmaetal2017}
{\sc \au{{Sharma}, R.}, \au{{Mitra}, D.} \& \au{{Oberoi}, D.}} \yr{2017}
  \at{{On the energization of charged particles by fast magnetic
  reconnection}}.  \jt{Monthly Notices of the Royal Astronomical Society}
  \bvol{470},  \pg{723--731}.

\bibitem[{Shibata} \& {Magara}(2011)]{shibata_solar_2011}
{\sc \au{{Shibata}, K.} \& \au{{Magara}, T.}} \yr{2011}  \at{{Solar Flares:
  Magnetohydrodynamic Processes}}.  \jt{Living Reviews in Solar Physics}
  \bvol{8},  \pg{6}.

\bibitem[{Sironi} \& {Spitkovsky}(2014)]{sironispitkovsky2014}
{\sc \au{{Sironi}, L.} \& \au{{Spitkovsky}, A.}} \yr{2014}  \at{{Relativistic
  Reconnection: An Efficient Source of Non-thermal Particles}}.
  \jt{Astrophysical Journal Letters}  \bvol{783},  \pg{L21}.

\bibitem[{Sweet}(1958)]{sweet_1958}
{\sc \au{{Sweet}, P.~A.}} \yr{1958} {The Neutral Point Theory of Solar Flares}.
   \bt{In {\em Electromagnetic Phenomena in Cosmical Physics\/} (ed.
  \ed{B.~{Lehnert}})},  \st{IAU Symposium},  \vol{vol.~6},  \pg{p. 123}.

\bibitem[{Tajima} \& {Shibata}(2002)]{tajimashibatabook}
{\sc \au{{Tajima}, T.} \& \au{{Shibata}, K.}} \yr{2002} {\em {Plasma
  astrophysics}\/}.  \publ{{Perseus, Cambridge, Mass.}}

\bibitem[{Tolman} {\em et~al.\/}(2018){Tolman}, {Loureiro} \&
  {Uzdensky}]{tolman2017}
{\sc \au{{Tolman}, E.~A.}, \au{{Loureiro}, N.~F.} \& \au{{Uzdensky}, D.~A.}}
  \yr{2018}  \at{{Development of tearing instability in a current sheet forming
  by sheared incompressible flow}}.  \jt{Journal of Plasma Physics}
  \bvol{84}~(1),  \pg{905840115}.

\bibitem[{Uzdensky}(2011)]{uzdensky2011}
{\sc \au{{Uzdensky}, D.~A.}} \yr{2011}  \at{{Magnetic Reconnection in Extreme
  Astrophysical Environments}}.  \jt{\ssr}  \bvol{160},  \pg{45--71}.

\bibitem[{Uzdensky} \& {Loureiro}(2016)]{uzdensky_2016}
{\sc \au{{Uzdensky}, D.~A.} \& \au{{Loureiro}, N.~F.}} \yr{2016}  \at{{Magnetic
  Reconnection Onset via Disruption of a Forming Current Sheet by the Tearing
  Instability}}.  \jt{Physical Review Letters}  \bvol{116}~(10),  \pg{105003}.

\bibitem[{Uzdensky} {\em et~al.\/}(2010){Uzdensky}, {Loureiro} \&
  {Schekochihin}]{uzdenskyetal2010}
{\sc \au{{Uzdensky}, D.~A.}, \au{{Loureiro}, N.~F.} \& \au{{Schekochihin},
  A.~A.}} \yr{2010}  \at{{Fast Magnetic Reconnection in the Plasmoid-Dominated
  Regime}}.  \jt{Physical Review Letters}  \bvol{105}~(23),  \pg{235002}.

\bibitem[{Waelbroeck}(1989)]{waelbroeck89}
{\sc \au{{Waelbroeck}, F.~L.}} \yr{1989}  \at{{Current sheets and nonlinear
  growth of the m=1 kink-tearing mode}}.  \jt{Physics of Fluids B}  \bvol{1},
  \pg{2372--2380}.

\bibitem[{Werner} \& {Uzdensky}(2017)]{wernUzd2017}
{\sc \au{{Werner}, G.~R.} \& \au{{Uzdensky}, D.~A.}} \yr{2017}  \at{{Nonthermal
  Particle Acceleration in 3D Relativistic Magnetic Reconnection in Pair
  Plasma}}.  \jt{Astrophysical Journal Letters}  \bvol{843},  \pg{L27}.

\bibitem[{Werner} {\em et~al.\/}(2016){Werner}, {Uzdensky}, {Cerutti},
  {Nalewajko} \& {Begelman}]{werner_2016}
{\sc \au{{Werner}, G.~R.}, \au{{Uzdensky}, D.~A.}, \au{{Cerutti}, B.},
  \au{{Nalewajko}, K.} \& \au{{Begelman}, M.~C.}} \yr{2016}  \at{{The Extent of
  Power-law Energy Spectra in Collisionless Relativistic Magnetic Reconnection
  in Pair Plasmas}}.  \jt{The Astrophysical Journal Letters}  \bvol{816},
  \pg{L8}.

\bibitem[{White} \& {Hazeltine}(2017)]{whitehazeltine17}
{\sc \au{{White}, R.~L.} \& \au{{Hazeltine}, R.~D.}} \yr{2017}  \at{{Analysis
  of the Hermite spectrum in plasma turbulence}}.  \jt{Physics of Plasmas}
  \bvol{24}~(10),  \pg{102315}.

\bibitem[{Yamada} {\em et~al.\/}(1997){Yamada}, {Ji}, {Hsu}, {Carter},
  {Kulsrud}, {Bretz}, {Jobes}, {Ono} \& {Perkins}]{yamadaetal1997}
{\sc \au{{Yamada}, M.}, \au{{Ji}, H.}, \au{{Hsu}, S.}, \au{{Carter}, T.},
  \au{{Kulsrud}, R.}, \au{{Bretz}, N.}, \au{{Jobes}, F.}, \au{{Ono}, Y.} \&
  \au{{Perkins}, F.}} \yr{1997}  \at{{Study of driven magnetic reconnection in
  a laboratory plasma}}.  \jt{Physics of Plasmas}  \bvol{4},  \pg{1936--1944}.

\bibitem[{Zhang} \& {Ji}(2014)]{zhangJi14}
{\sc \au{{Zhang}, Q.~M.} \& \au{{Ji}, H.~S.}} \yr{2014}  \at{{Blobs in
  recurring extreme-ultraviolet jets}}.  \jt{Astronomy and Astrophysics}
  \bvol{567},  \pg{A11}.

\bibitem[{Zhang} {\em et~al.\/}(2012){Zhang}, {Lu}, {Baumjohann}, {Russell},
  {Fedorov}, {Barabash}, {Coates}, {Du}, {Cao}, {Nakamura}, {Teh}, {Wang},
  {Dou}, {Wang}, {Glassmeier}, {Auster} \& {Balikhin}]{zhangetal12}
{\sc \au{{Zhang}, T.~L.}, \au{{Lu}, Q.~M.}, \au{{Baumjohann}, W.},
  \au{{Russell}, C.~T.}, \au{{Fedorov}, A.}, \au{{Barabash}, S.}, \au{{Coates},
  A.~J.}, \au{{Du}, A.~M.}, \au{{Cao}, J.~B.}, \au{{Nakamura}, R.}, \au{{Teh},
  W.~L.}, \au{{Wang}, R.~S.}, \au{{Dou}, X.~K.}, \au{{Wang}, S.},
  \au{{Glassmeier}, K.~H.}, \au{{Auster}, H.~U.} \& \au{{Balikhin}, M.}}
  \yr{2012}  \at{{Magnetic Reconnection in the Near Venusian Magnetotail}}.
  \jt{Science}  \bvol{336},  \pg{567}.

\bibitem[{Zhdankin} {\em et~al.\/}(2013){Zhdankin}, {Uzdensky}, {Perez} \&
  {Boldyrev}]{zhdankinetal2013}
{\sc \au{{Zhdankin}, V.}, \au{{Uzdensky}, D.~A.}, \au{{Perez}, J.~C.} \&
  \au{{Boldyrev}, S.}} \yr{2013}  \at{{Statistical Analysis of Current Sheets
  in Three-dimensional Magnetohydrodynamic Turbulence}}.  \jt{Astrophysical
  Journal}  \bvol{771},  \pg{124}.

\bibitem[{Zhou} {\em et~al.\/}(2015){Zhou}, {B{\"u}chner}, {B{\'a}rta}, {Gan}
  \& {Liu}]{zhouetal2015}
{\sc \au{{Zhou}, X.}, \au{{B{\"u}chner}, J.}, \au{{B{\'a}rta}, M.}, \au{{Gan},
  W.} \& \au{{Liu}, S.}} \yr{2015}  \at{{Electron Acceleration by Cascading
  Reconnection in the Solar Corona. I. Magnetic Gradient and Curvature Drift
  Effects}}.  \jt{Astrophysical Journal}  \bvol{815},  \pg{6}.

\bibitem[{Zocco} {\em et~al.\/}(2015){Zocco}, {Loureiro}, {Dickinson}, {Numata}
  \& {Roach}]{zocco_2015}
{\sc \au{{Zocco}, A.}, \au{{Loureiro}, N.~F.}, \au{{Dickinson}, D.},
  \au{{Numata}, R.} \& \au{{Roach}, C.~M.}} \yr{2015}  \at{{Kinetic
  microtearing modes and reconnecting modes in strongly magnetised slab
  plasmas}}.  \jt{Plasma Physics and Controlled Fusion}  \bvol{57}~(6),
  \pg{065008}.

\bibitem[{Zocco} \& {Schekochihin}(2011)]{zs2011}
{\sc \au{{Zocco}, A.} \& \au{{Schekochihin}, A.~A.}} \yr{2011}  \at{{Reduced
  fluid-kinetic equations for low-frequency dynamics, magnetic reconnection,
  and electron heating in low-beta plasmas}}.  \jt{Physics of Plasmas}
  \bvol{18}~(10),  \pg{102309--102309}.

\bibitem[{Zong} {\em et~al.\/}(2004){Zong}, {Fritz}, {Pu}, {Fu}, {Baker},
  {Zhang}, {Lui}, {Vogiatzis}, {Glassmeier}, {Korth}, {Daly}, {Balogh} \&
  {Reme}]{zongetal2004}
{\sc \au{{Zong}, Q.-G.}, \au{{Fritz}, T.~A.}, \au{{Pu}, Z.~Y.}, \au{{Fu},
  S.~Y.}, \au{{Baker}, D.~N.}, \au{{Zhang}, H.}, \au{{Lui}, A.~T.},
  \au{{Vogiatzis}, I.}, \au{{Glassmeier}, K.-H.}, \au{{Korth}, A.}, \au{{Daly},
  P.~W.}, \au{{Balogh}, A.} \& \au{{Reme}, H.}} \yr{2004}  \at{{Cluster
  observations of earthward flowing plasmoid in the tail}}.  \jt{Geophysical
  Review Letters}  \bvol{31},  \pg{L18803}.

\bibitem[{Zweibel} \& {Yamada}(2009)]{ZY2009}
{\sc \au{{Zweibel}, E.~G.} \& \au{{Yamada}, M.}} \yr{2009}  \at{{Magnetic
  Reconnection in Astrophysical and Laboratory Plasmas}}.  \jt{Annual Review of
  Astronomy and Astrophysics}  \bvol{47},  \pg{291--332}.

\end{thebibliography}

\end{document}